\documentclass[journal,draftclsnofoot,onecolumn]{IEEEtran}

\usepackage{amsmath}
\allowdisplaybreaks[1]
\usepackage{setspace}

\usepackage{amssymb}
\usepackage{mathrsfs}
\usepackage{amsthm}
\usepackage{multirow}
\usepackage{color}
\usepackage{longtable}
\usepackage{array}
\usepackage{url}
\usepackage{comment}
\usepackage{enumerate}
\usepackage{eucal}
\usepackage{graphics}
\usepackage{verbatim}
\usepackage{subfigure}

\usepackage{algorithm}
\usepackage{algpseudocode}
\usepackage{cite}
\usepackage[table]{xcolor}

\usepackage{mathtools}
\usepackage{xparse}
\usepackage{bm}

\DeclarePairedDelimiterX{\set}[1]{\{}{\}}{\setargs{#1}}
\NewDocumentCommand{\setargs}{>{\SplitArgument{1}{;}}m}
{\setargsaux#1}
\NewDocumentCommand{\setargsaux}{mm}
{\IfNoValueTF{#2}{#1} {#1\,\delimsize|\,\mathopen{}#2}}

\newtheorem{thm}{Theorem}

\newtheorem{lem}{Lemma}

\newtheorem{prop}{Proposition}
\newtheorem{rmk}{Remark}
\newtheorem{cor}{Corollary}
\newtheorem{example}{Example}
\newenvironment{pf}{{\noindent\it Proof:}}{\hfill $\blacksquare$\par}

\newcommand{\RNum}[1]{\lowercase\expandafter{\romannumeral #1\relax}}
\newcommand{\Rnum}[1]{\uppercase\expandafter{\romannumeral #1\relax}}

\newcommand{\F}{\mathbb{F}}

\newcommand{\eqdef}{\triangleq}
\newcommand{\cA}{\mathcal{A}}
\newcommand{\cB}{\mathcal{B}}

\newcommand{\cD}{\mathcal{D}}
\newcommand{\cE}{\mathcal{E}}

\newcommand{\cN}{\mathcal{N}}
\newcommand{\cO}{\mathcal{O}}

\newcommand{\cU}{\mathcal{U}}
\newcommand{\cV}{\mathcal{V}}
\newcommand{\cW}{\mathcal{W}}
\newcommand{\cT}{\mathcal{T}}

\newcommand{\cS}{\mathcal{S}}
\newcommand{\cH}{\mathcal{H}}

\newcommand{\bb}{\mathbf{b}}
\newcommand{\bc}{\mathbf{c}}
\newcommand{\bd}{\mathbf{d}}

\newcommand{\bu}{\mathbf{u}}

\newcommand{\bx}{\mathbf{x}}

\newcommand{\Zero}{\mathbf{0}}

\newcommand{\In}{{\rm In}}
\newcommand{\Out}{{\rm Out}}
\newcommand{\tail}{{\rm tail}}
\newcommand{\head}{{\rm head}}

\newcommand{\Rank}{{\rm Rank}}
\newcommand{\Mod}{{\rm Mod}}
\newcommand{\mincut}{{\rm min\textup{-}cut }}

\begin{document}

\title{On hierarchical secure aggregation against relay and user collusion}

\author{
  Min Xu, Xuejiao Han, Kai Wan and  Gennian Ge%
  \thanks{This research was supported by the National Key Research and Development Program of China under Grant 2025YFC3409900, the National Natural Science Foundation of China under Grant 12231014, and Beijing Scholars Program.}
  \thanks{M. Xu (e-mail: minxu0716@qq.com) is with the Institute of Mathematics and Interdisciplinary Sciences, Xidian University, Xi'an 710126, China.}%
  \thanks{X. Han (e-mail: hanxj4425@126.com) and G. Ge (e-mail: gnge@zju.edu.cn) are with the School of Mathematical Sciences, Capital Normal University, Beijing 100048, China.}
  \thanks{K.~Wan (e-mail: kai\_wan@hust.edu.cn) is with the School of Electronic Information and Communications, Huazhong University of Science and Technology, 430074 Wuhan, China.}
}

\maketitle
\begin{abstract}
Secure aggregation (SA) is fundamental to privacy preservation in federated learning (FL), enabling model aggregation while preventing disclosure of individual user updates. 
This paper addresses hierarchical secure aggregation (HSA) against relay and user collusion in homogeneous networks, where each user connects to $n$ relays and each relay serves $m$ users. 
In the two-phase communication framework, users transmit masked data to relays, which then process and forward compiled messages to the server for exact sum recovery. 
The primary objective is to devise a transmission scheme such that the server can finish the aggregation task, while any group of $T_h$ colluding relays and $T_u$ colluding users cannot reveal any information about the data owned by the non-colluding users. 
In this study, we establish fundamental limits on the communication load, defined as the ratio of transmitted information size to original data size, for each user-relay link and each relay-server link.
Achievable thresholds for collusion resilience are also derived. When the number of colluding relays and users falls below certain critical thresholds, we construct communication-optimal schemes using methods from network function computation. A limitation of these schemes is their reliance on large random keys. To address this, we derive a lower bound on the required key size and prove its achievability in cyclic networks, where users are connected to relays in a cyclic wrap-around manner.  By establishing a connection between HSA and network function computation, this work advances the theoretical limits of communication efficiency and information-theoretic security in secure aggregation.
\end{abstract}
\begin{IEEEkeywords}
Secure aggregation, hierarchical network, federated learning, security, relay-user collusion
\end{IEEEkeywords}

\section{Introduction}\label{sec:introduction}

Secure aggregation has emerged as a critical research domain \cite{2017bonawitzpractical,2022ZhaoSun}, driven by the increasing importance of data security and user privacy in the context of federated learning (FL) \cite{2017mcmahancommunication,2016konevcnyfederated,2021kairouzadvances}. The main challenge in this context is to enable a central server to compute an aggregated model from locally trained user models without gaining access to any individual user’s private data. To address this challenge, a variety of secure aggregation protocols have been proposed \cite{2017bonawitzpractical,2022ZhaoSun,2020WeiLiDing,2020HuGuoLi,2021ZhaoZhaoYang,2021andrewdifferentially,2024YeminiSaha,2024lindifferential,2021SoGuler,2020kadhefastsecagg,2022ElkordyAvestimehr,2023LiuGuoLam,2023JahaniMALi}, many of which employ cryptographic techniques to ensure security. A seminal contribution to this field was the secure aggregation protocol introduced by Bonawitz {\it et al.} \cite{2017bonawitzpractical}. This protocol utilizes a pairwise random seed agreement mechanism, in which users establish mutual agreements to generate random keys. Upon aggregation, the collective effect of these random keys neutralizes, enabling the retrieval of the summation of the local models, while concealing other information about the users' models. Additionally, the protocol integrates Shamir’s secret sharing \cite{shamir1979share} to enable key recovery, thereby ensuring robustness against user dropouts or server-user collusion. To improve efficiency, So {\it et al.} \cite{2021SoGuler} introduced a secure aggregation protocol that reduces the quadratic overhead inherent in the pairwise random seed agreement scheme of \cite{2017bonawitzpractical}. Alternative paradigms for secure aggregation have explored approaches such as multi-secret sharing \cite{2020kadhefastsecagg}, secure multi-party computation (MPC) \cite{2022ElkordyAvestimehr}, and polynomial interpolation \cite{2023JahaniMALi}. A significant limitation in the random seed-based key generation schemes lies in their inability to achieve information-theoretic security, as constrained by Shannon's one-time pad theorem \cite{shannon1949communication}. In contrast, differential privacy (DP) \cite{2021andrewdifferentially,2020HuGuoLi,2024lindifferential,2020WeiLiDing} presents an alternative approach, employing the introduction of small random perturbations to protect local models. A critical drawback of this approach is that the individual noise components do not completely cancel out during the summation process, which consequently introduces imprecision into the aggregate model. The research presented in \cite{2020WeiLiDing} demonstrated a discernible trade-off between the level of privacy protection (characterized by noise strength) and the overall convergence speed of the model. Despite their operational simplicity, DP-based techniques inherently provide only probabilistic privacy guarantees, rather than perfect privacy.

Information-theoretic secure aggregation (ITSA) provides the strongest security guarantees, offering perfect privacy through the assurance of zero mutual information between the server and the individual user data. The primary objective of ITSA revolves around optimizing both the communication overhead associated with user uploads and the computational efficiency of secret key generation. Numerous studies have delved into these objectives across a variety of application scenarios \cite{2022ZhaoSun,2023JahaniMALi,2021SoGuler,2022JahaniMALi,2024Zhaosun,2024WanYaoSunJiCaire,2024WanSunJiCaire,2023ZhaoSunISIT,2025ZhaoSunTIT,2023LiZhaoSun,2023sunsecure,2025yuanSun,2021karakoccsecure}. For example, previous works, such as \cite{2022ZhaoSun,2023JahaniMALi,2021SoGuler,2022JahaniMALi}, investigated solutions to maintain security and data integrity in face of user dropout (the unavailability of inputs from a subset of users) and collusion scenarios (where the server may collude with certain users). The utilization of groupwise keys \cite{2024Zhaosun,2024WanSunJiCaire,2024WanYaoSunJiCaire} provided efficient frameworks where each key is shared by a subset of users, and the keys are mutually independent. Moreover, researches such as \cite{2025ZhaoSunTIT,2023ZhaoSunISIT,2023sunsecure,2021karakoccsecure}, specifically addressed secure aggregation protocols that incorporate strategies such as proactive user selection, operation under an oblivious server model, and consideration of the threats posed by malicious participating users.

A related area of active research concerns the impact of network topology on the performance of secure aggregation. Hierarchical secure aggregation (HSA) has been studied under various network models. Among them, a widely adopted model is the tree-structured network, where each user is connected to exactly one relay. This model was considered in \cite{2024zhangwansunwangoptimal,2024ZhangWanSunWangITW}, and also in \cite{2025LiZhangLvFan} under a weakly security constraint.
To capture more flexible device associations, several works have explored generalized topologies. Egger {\it et al.} \cite{2023EggerHofmeister,egger2023private} considered a model in which a user may connect to multiple relays. Their approach involves a secret key aggregation phase that requires sequential communication among relays, introducing additional latency. Lu {\it et al.} \cite{2024Luchengkangliucapacity} studied a fully-connected topology where every user is linked to every relay. While this model also entails inter-relay communication to cancel keys, the associated communication cost was not explicitly characterized.
Further extending architectural flexibility, Zhang {\it et al.} \cite{2025HSAcyclic} introduced a cyclic association model that supports arbitrary link patterns between users and relays. By employing correlated key generation, their design avoids inter-relay communication altogether. 
A number of works have also addressed the issue of user collusion, though mainly within simplified network settings \cite{2024zhangwansunwangoptimal,2024ZhangWanSunWangITW,2025LiZhangLvFan}. While these efforts offered valuable insights, the analysis of collusion resilience in general network topologies remains an open problem.

In this paper, we significantly advance the study of HSA by explicitly incorporating the problem of \emph{relay and user collusion} within a generalized and practically relevant homogeneous user-relay association model. We consider a network consisting of $N$ users and $K$ relays, where each user is connected to exactly $n$ relays, and each relay is assigned to receive messages from exactly $m$ users. This model generalizes the restrictive tree assumption, resembling a cyclic association structure, and supports a more distributed and fault-tolerant relaying mechanism. The system operates in two phases: first, each user securely transmits its input to all its associated relays; then, each relay further encodes the received partial messages and forwards an encrypted aggregated message to the central server. The server aims to recover the sum of all users’ inputs while adhering to the security constraints, as established in \cite{2024zhangwansunwangoptimal}.

To characterize potential adversarial behaviors, we introduce a collusion model defined by the triple $(\mathcal{T}_s, \mathcal{T}_h, \mathcal{T}_u)$, where $\mathcal{T}_s\in \{\emptyset, \text{server}\}$ indicates whether the server colludes, $\mathcal{T}_h \subseteq [K]$ denotes a set of colluding relays, and $\mathcal{T}_u \subseteq [N]$ represents a set of colluding users. A scheme is said to achieve $(T_s, T_h, T_u)$-security if it can resist any coalition of up to $T_h$ relays, any group of up to $T_u$ users, and the server when $T_s = 1$. 
The primary goal is to design a secure aggregation scheme under this generalized collusion model and characterize the optimal rate region \(\mathcal{R}^*\), defined as the closure of all achievable tuples \((R_X, R_Y, R_Z, R_{Z_\Sigma})\), where \(R_X\) denotes the maximum communication load between all user-relay links, \(R_Y\) denotes the maximum communication load among all relay-server links, \(R_Z\) represents the key size stored by each user, and \(R_{Z_\Sigma}\) indicates the total key size across the entire system, with all parameters normalized by the size of the input data.
By analyzing collusion under this generalized model, we aim to provide practical insights for designing efficient and secure distributed aggregation systems.

\begin{figure}
    \subfigure[An example of HSA model]{\begin{minipage}[b]{0.49\linewidth}\includegraphics[width=1\linewidth]{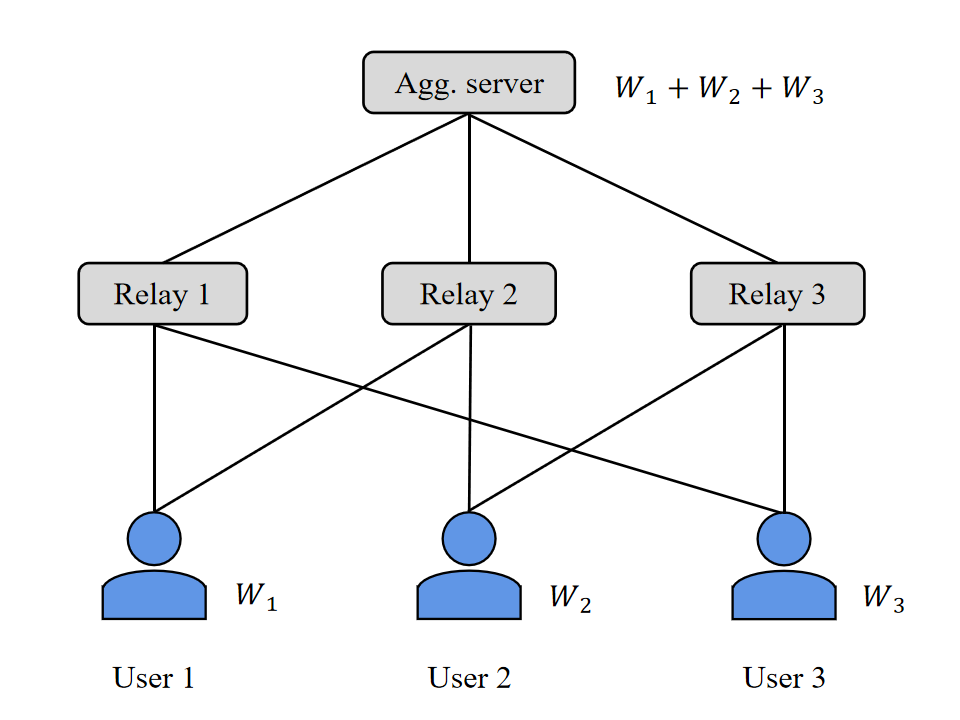}\end{minipage}}
    \vspace{4pt}
    \subfigure[The corresponding NFC model]{\begin{minipage}[b]{0.49\linewidth}\includegraphics[width=1\linewidth]{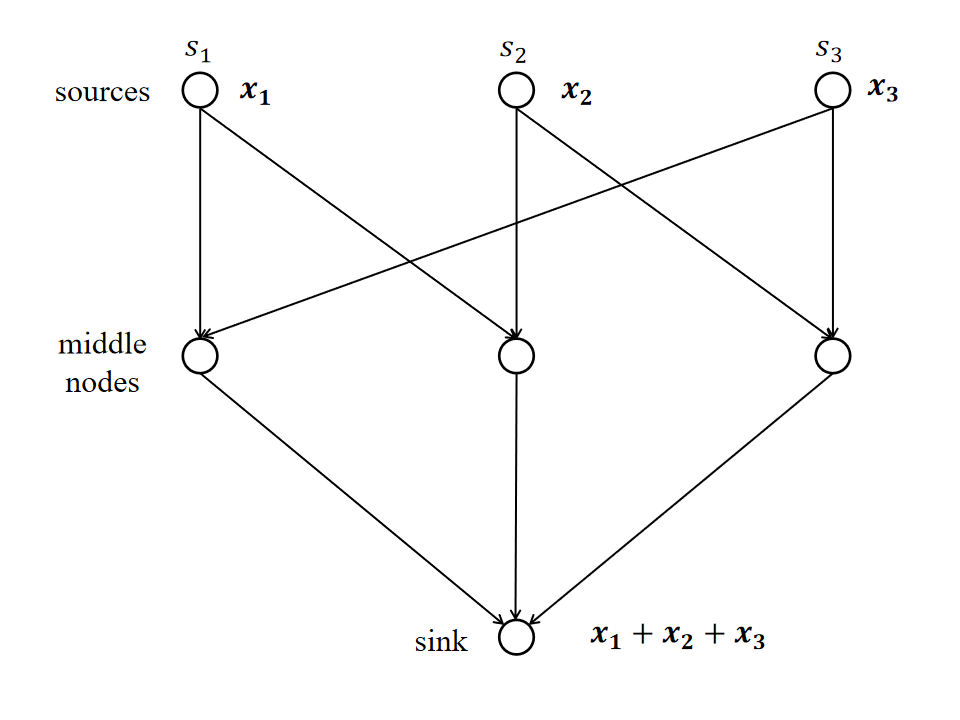}\end{minipage}}
    \caption{An illustration of the equivalence between HSA and NFC.}
    \label{fig:HSA2NFC}
\end{figure}

Another active research area in network information theory is \emph{network function computation} (NFC). In this problem, a sink node is required to compute a target function $f$ of source messages which are generated at multiple source nodes, and every intermediate node  can encode the received messages and transmit the result to the downstream nodes. The \emph{computing rate} of a network code is the average times that the target function $f$ can be  computed with zero error for one use of the network. The maximum computing rate is called the \emph{computing capacity}. The primary objective is to determine the computing capacity for given network topologies and target functions \cite{2011AFKZ,2018HTYG,2019GYYL}. 
Notably, the HSA problem can be regarded as a special case within this framework. 
Specifically, the users act as sources, the relays serve as middle nodes, and the function to be computed is the algebraic sum\footnote{We use the term ``algebraic sum'' to emphasize summation over finite fields, as opposed to arithmetic summation over rational numbers.} over a three-layer network (see Fig.~\ref{fig:HSA2NFC} as an illustration).
As a result, the communication load pair $(R_X, R_Y)$ in HSA corresponds directly to the computing capacity in the network function computation context. Prior works have developed capacity-achieving schemes for tree networks or for algebraic sums over finite fields \cite{2011AFKZ,2012RD}, and have established upper bounds for more general topologies and functions \cite{2011AFKZ,2018HTYG,2019GYYL}. However, determining the exact computing capacity for arbitrary functions and network topologies remains a main challenge--even for linear target functions \cite{2014Appuswamy}.
Additional requirements such as robustness and security have also been studied in network function computation \cite{2002CaiYeung,2011Cai,2010CaiYeung,2005Bhattad,2008Harada,2021GBY,2024Guangsourcesecure,2025BaiGuangFunctionsecure,WeiXuGe23}. The existing researches have primarily focused on \emph{link-security}, where eavesdroppers compromise specific edges in the network \cite{2002CaiYeung,2011Cai,2010CaiYeung,2005Bhattad,2008Harada,2021GBY,2024Guangsourcesecure,2025BaiGuangFunctionsecure}.
In this paper, the $(T_s,T_h,T_u)$-security notion corresponds to \emph{node-security} in network function computation, which has not been investigated until now. Analyzing node security for general network topologies is highly challenging due to the complexity of potential collusion patterns. This difficulty motivates our focus on a structured three-layer network, which provides a tractable but meaningful model for studying node-level security.

This paper focuses on the case where $T_s = 0$, i.e., the server is trusted and does not collude. The main contributions are summarized as follows.
\begin{enumerate}
    \item We establish a fundamental connection between HSA and NFC. Based on this connection, we construct a $(0,T_h,T_u)$-secure HSA scheme that achieves the communication load pair $(R_X,R_Y)=(1/n,1/n)$ for any homogeneous three-layer network. Notably, this result is shown to be optimal in terms of the cut-set bound. 
    \item We derive lower bounds on the key rates $R_Z$ and $R_{Z_\Sigma}$ using standard information-theoretic approaches akin to those in prior works \cite{2025HSAcyclic,2024zhangwansunwangoptimal}. However, these initial bounds are often not tight under complex network topologies and user-relay collusion constraints. To address this, we subsequently develop a refined lower bound for a special class of networks by incorporating decodability constraints. This improved bound is shown to be tight, as it matches the performance of our constructed scheme.
\end{enumerate}

The remainder of this paper is organized as follows. Section~\ref{sec:preliminary} introduces the system model and previous results of HSA and NFC. Section~\ref{sec:mainresults} presents our main theorems. The converse proofs for communication and key rate bounds are established in Section~\ref{sec:converse}, while Section~\ref{sec:scheme} details our constructive schemes. Section~\ref{sec:example} presents an example in which the lower bound for key rates is not strictly tight. Furthermore, we improve the lower bound for cyclic networks with particular parameters. Conclusions and future research directions are discussed in Section~\ref{sec:conclusion}.

\emph{Notation.} Throughout this paper, the following notations are used: Matrices are denoted by bold uppercase letters. Random variables and random vectors are denoted by uppercase letters. Deterministic Vectors (non-random) are denoted by boldface, lowercase letters. For integers $m\leq n$, let $[m:n] \eqdef \{m, m + 1, \cdots , n\}$ if $m \leq n$ and
$[m : n] = \emptyset $, if $m > n$. $[1 : n]$ is written as $[n]$ for brevity. $\Mod(a, b)$
denotes the remainder of $a$ divided by $b$.

\section{System model and related results}\label{sec:preliminary}

In this section, we present the system model and relevant results in HSA. As NFC serves as a key methodology in this work, we also describe its basic model and establish the relationship between HSA and NFC. Additionally, we introduce some notations and results in NFC that will be utilized in the construction of the proposed scheme.
\subsection{System model}
\begin{figure}
    \centering
    \includegraphics[width=0.6\linewidth]{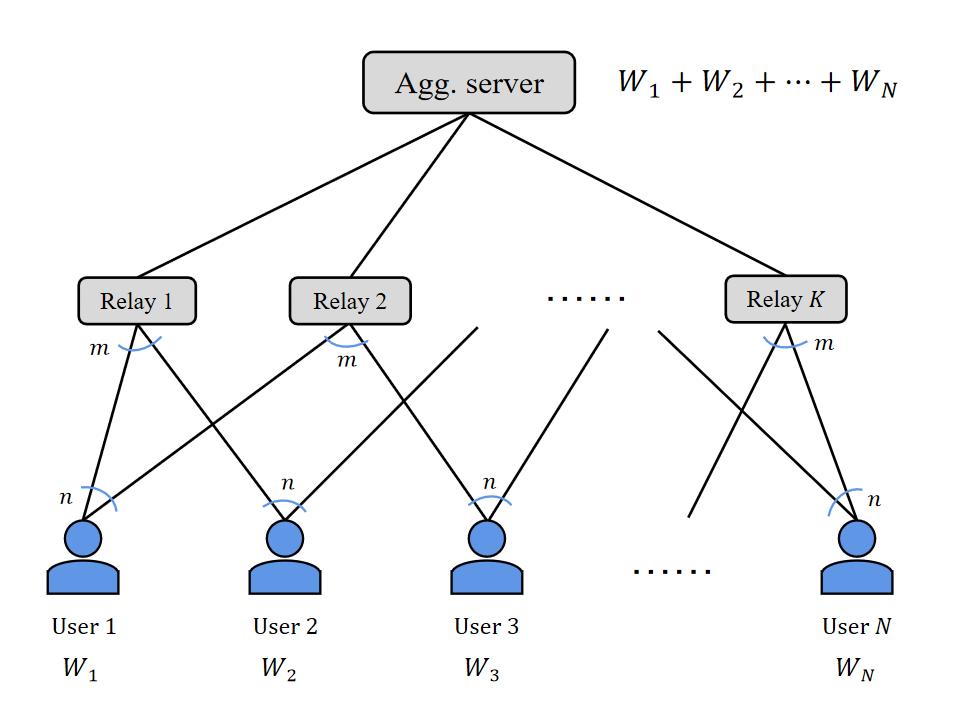}
    \caption{A homogeneous three-layer network $\cN_{K,K,n}$.}
    \label{fig:model}
\end{figure}

This paper studies the HSA problem in a homogeneous three-layer network $\mathcal{N}_{N,K,n}$, as illustrated in Fig.~\ref{fig:model}. The network consists of $N$ users, $K$ relays, and one aggregation server. Each user is connected to $n$ relays, and each relay is associated with $m$ users, such that $Nn = Km$. We assume $n < K$ throughout the paper. This type of network topology is commonly found in distributed coded computing~\cite{2021CDCsurvey,2018Lisongze} and gradient coding schemes~\cite{2017tandongradient,2019blockdesign}, where users can be interpreted as data sources and relays as processing units.
Let $\mathcal{H}_i$ denote the set of relays connected to the $i$-th user, and $\mathcal{U}_j$ denote the set of users connected to the $j$-th relay. Each user $i$ holds an input $W_i$, which contains $L$ symbols from a finite field $\mathbb{F}_q$. The user inputs are assumed to be selected independently and identically distributed (i.i.d.) and uniformly at random. Thus, the entropy of each input is $H(W_i) = L$ (in $q$-ary units), and the total entropy of all inputs is
\[
H(W_1, \cdots, W_N) = \sum_{i=1}^{N} H(W_i) = NL.
\]

To ensure security, each user $i$ is also equipped with a secret key $Z_i$ comprising $L_Z$ symbols. Each $Z_i$ is known only to user $i$ and is kept private from the relays and the aggregation server. These individual keys are derived from a source key $Z_{\Sigma}$ containing $L_{Z_{\Sigma}} = H(Z_1, \cdots, Z_N)$ symbols, such that
\[
H(Z_1, \cdots, Z_N \mid Z_{\Sigma}) = 0.
\]
A trusted third party generates and distributes the individual keys to prevent information leakage to the server or relays. Moreover, the keys $Z_1, \cdots, Z_N$ are independent of the user inputs, satisfying
\[
H(W_{[N]}, Z_{[N]}) = \sum_{i=1}^{N} H(W_i) + H(Z_{[N]}).
\]
The goal of the aggregation server is to compute the sum of all user inputs, $\sum_{i=1}^{N} W_i$.

The security constraint considered in this study is $(0,T_h,T_u)$-security, which requires that any collusion of up to $T_h$ relays and $T_u$ users must not be able to infer information about the complete set of user inputs $W_{[N]}$.
In the first hop, each user $i \in [N]$ sends a message $X_{i,j}$ to each connected relay $j \in \mathcal{H}_i$. The messages transmitted by each user depend solely on its own input and secret key, i.e.,
\begin{align}
    \label{eq:first_hop_determinism}
    H(\{X_{i,j}\}_{j \in \mathcal{H}_i} | W_i, Z_i) = 0, \quad \forall i \in [N].
\end{align}
In the second hop, each relay $j \in [K]$ collects all messages from its connected users, $\{X_{i,j}\}_{i \in \mathcal{U}_j}$, and transmits an encoded message $Y_j$ to the server. Thus,
\begin{align}
    \label{eq:second_hop_determinism}
    H(Y_j | \{X_{i,j}\}_{i \in \mathcal{U}_j}) = 0, \quad \forall j \in [K].
\end{align}
For two integers $T_h\geq1,T_u\geq0$, a hierarchical secure aggregation scheme is  \emph{$(0,T_h,T_u)$-secure} if it satisfies the following two conditions:
\begin{itemize}
    \item \emph{Decodability}. The server should be able to perfectly reconstruct the sum of all user inputs from the signals received from the relays:
    \begin{equation}
        \label{eq:decodability}
        H\left(\sum_{i=1}^{N} W_i \Big| \{Y_j\}_{j \in [K]}\right) = 0.
    \end{equation}

    \item \emph{$(0,T_h,T_u)$-Security}. Any group of at most $T_h$ relays and $T_u$ users must not gain any information about the full input set $W_{[N]}$. Formally, for all $\mathcal{T}_h \subset [K]$ with $|\mathcal{T}_h| \le T_h$ and $\mathcal{T}_u \subset [N]$ with $|\mathcal{T}_u| \le T_u$,
    \begin{equation}
        \label{eq:security}
        I(W_{[N]}; \{X_{i,j} : i \in \mathcal{U}_j\}_{j \in \mathcal{T}_h} \mid \{W_{i'}, Z_{i'}\}_{i' \in \mathcal{T}_u}) = 0.
    \end{equation}
\end{itemize}

We refer to the class of problems defined by the network $\mathcal{N}_{N,K,n}$ and security parameters $T_u, T_h$ as the hierarchical secure aggregation problem, formally represented by the tuple $(N,K,n,T_u,T_h,0)$. Here, the value $0$ specifies that the server is not part of the collusion set.
This paper investigates both the communication and key generation of the hierarchical secure aggregation problem. 
We define the user-to-relay communication load, $R_X$, and the relay-to-server communication load, $R_Y$, as:
\[R_X \triangleq \max_{i\in[N],j\in\cH_i} \frac{H(X_{i,j})}{L}, \quad R_Y \triangleq \max_{j\in[K]}\frac{ H(Y_j)}{L}.\]
Similarly, we define the individual key rate, $R_Z$, and the source key rate, $R_{Z_{\Sigma}}$, as:
\[R_Z \triangleq \frac{L_Z}{L}, \quad R_{Z_\Sigma} \triangleq \frac{L_{Z_\Sigma}}{L}.\]
The rates $R_X$ and $R_Y$ reflect the communication efficiency, while $R_Z$ and $R_{Z_\Sigma}$ reflect the efficiency of key generation in the secure aggregation scheme. A rate tuple $(R_X, R_Y, R_Z, R_{Z_\Sigma})$ is considered achievable if there exists a $(0, T_h, T_u)$-secure HSA scheme that attains these rates. The optimal rate region $\mathcal{R}^*$ is defined as the closure of all achievable rate tuples.

Previous studies on secure aggregation \cite{2022ZhaoSun,2023JahaniMALi,2021SoGuler,2022JahaniMALi,2024Zhaosun,2024WanYaoSunJiCaire,2024WanSunJiCaire,2023ZhaoSunISIT,2025ZhaoSunTIT,2023LiZhaoSun,2023sunsecure,2025yuanSun,2021karakoccsecure,2024zhangwansunwangoptimal,2025HSAcyclic} have primarily focused on two security paradigms: relay security and server security, corresponding respectively to the notions of $(0,1,T)$-security and $(1,0,T)$-security introduced in this paper. The latter captures settings in which the server colludes with a subset of users. This work, however, does not address such server collusion scenarios.
Among existing literature, the most relevant to our work are \cite{2024zhangwansunwangoptimal,2025HSAcyclic,2025LiZhangLvFan}. In \cite{2024zhangwansunwangoptimal}, Zhang {\it et al.} studied the HSA problem under user collusion in tree-structured networks. In this network model, there are $U$ relays, each associated with $V$ users (that is, $N=UV, K=U$ and $n=1$). The following rate region was established:
\begin{thm}[Theorem~1 \cite{2024zhangwansunwangoptimal}]\label{thm:treewithcolluding}
    For the hierarchical secure aggregation problem with $U \geq 2$ relays, $V$ users per cluster and a maximum of $T$ colluding users, the optimal rate region is given by
    \[\mathcal{R}^* = \begin{cases} 
        \left\{R_X \geq 1,  R_Y \geq 1, R_Z \geq 1,  
        R_{Z_\Sigma} \geq \max\{V+T, \min\{UV - 1, U + T - 1\}\}\right\}, & \text{if } T < (U - 1)V; \\ 
        \emptyset, & \text{if } T \geq (U - 1)V. 
    \end{cases}\]
\end{thm}
It should be noted that $R_{Z_\Sigma} \geq \min\{UV - 1, U + T - 1\}$ is based on the server security assumptions in \cite{2024zhangwansunwangoptimal}.
In the very recent study \cite{2025LiZhangLvFan}, Li {\it et al.} extended both the tree network model and the security constraints to a heterogeneous setting. Specifically, each relay in the tree network may be connected to a different number of users. Moreover, the collusion set and the secure input set are characterized by two predefined families $\bm{\mathcal{T}}$ and $\bm{\mathcal{S}}$. In our model, when $T_h=1$, the $(0,1,T)$-security and $(1,0,T)$-security correspond to $\bm{\mathcal{T}} = \{\mathcal{T} \subseteq [N] : |\mathcal{T}| \leq T\}$ and $\bm{\mathcal{S}} = \{\mathcal{S} : \mathcal{S} \subseteq [N]\}$. The authors addressed the problem by dividing it into three cases, establishing a lower bound on $R_{Z_\Sigma}$ for each. They also proposed schemes satisfying the security constraints, with those for the first two cases shown to achieve their corresponding bounds. As the explicit form of the bound involves substantial notation that is not used in this paper, we omit the details of their results here.

\begin{figure}
    \subfigure[An example of 3-layer network]{\begin{minipage}[b]{0.49\linewidth}\includegraphics[width=1\linewidth]{HSA.png}\end{minipage}}
    \vspace{4pt}
    \subfigure[The corresponding tree network]{\begin{minipage}[b]{0.49\linewidth}\includegraphics[width=1\linewidth]{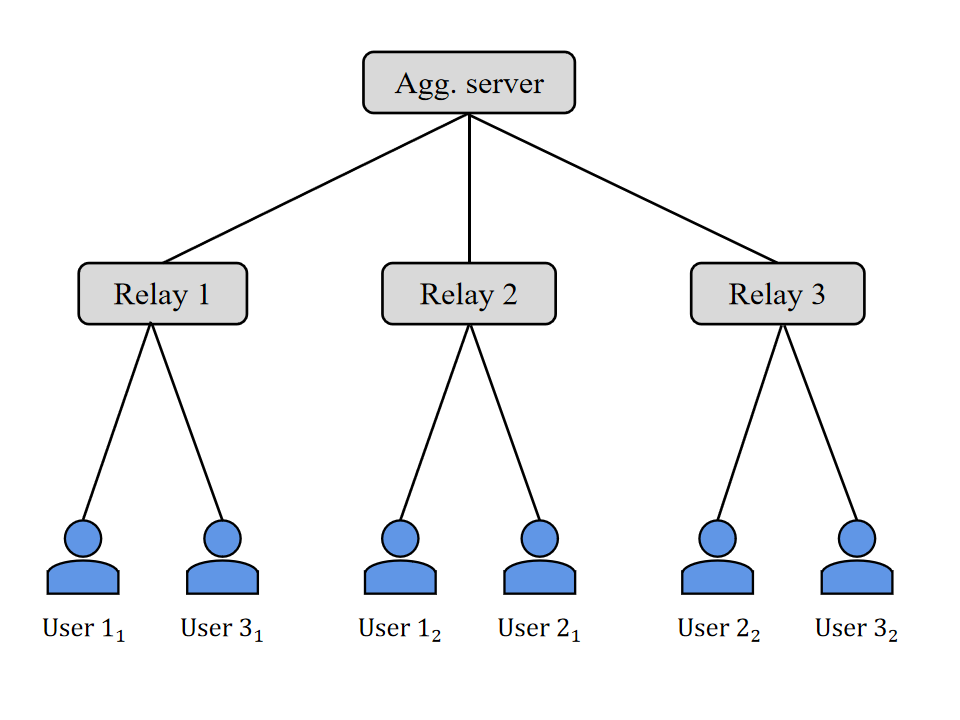}\end{minipage}}
    \caption{An illustration of transforming a 3-layer network to a tree.}
    \label{fig:HSA2Tree}
\end{figure}

\begin{rmk}

    For the homogeneous three-layer network model considered in this paper, each user $i$ is connected to $n$ relays, denoted by $\mathcal{H}_i$. We define an operation called \emph{replacing user $i$ with $n$ virtual users $i_1, \cdots, i_n$} as follows: (i) removing user $i$ and introducing $n$ virtual users $i_1, \cdots, i_n$; (ii) connecting each virtual user $i_j$ to exactly one relay in $\mathcal{H}_i$, such that every relay in $\mathcal{H}_i$ is associated with exactly one virtual user. By applying this operation to every user, the original homogeneous three-layer network $\mathcal{N}_{N,K,n}$ is transformed into a tree network with $Nn$ virtual users and $K$ relays (an example is provided in Fig.~\ref{fig:HSA2Tree}).

    As a result, the $(0,1,T)$-security notion introduced in this paper corresponds to a collusion set comprising $nT$ virtual users in the transformed tree network. This establishes a conceptual alignment with the model studied in \cite{2025LiZhangLvFan}, though two important distinctions must be noted. 
    First, to achieve optimal communication load, each virtual user in our model possesses only a $1/n$ fraction of the original user's data. Therefore, the goal of the server is not to compute the simple sum of all virtual users' values, but a vector linear function of the messages held by these virtual users. 
    Second, Lemma 4 in \cite{2025LiZhangLvFan} relies on the assumption that removing any user $(u',v')$ from a collusion set $\mathcal{T}_n$ preserves the collusion property. This assumption does not hold in our setting. For example, when $T = 1$, the collusion set consists of the $n$ virtual users derived from a single original user. Removing any one virtual user invalidates the collusion condition, since all virtual users originate from the same physical user. Due to this structural difference, the results of \cite{2025LiZhangLvFan} are not directly applicable to our model, even in the case where $T_h = 1$.
\end{rmk}

To explore the impact of network topology on the performance of HSA, Zhang {\it et al.} \cite{2025HSAcyclic} extended the network model by considering a cyclic association structure and examined the case without user collusion. They established the following optimal rate region.
\begin{thm}[Theorem~1 \cite{2025HSAcyclic}]\label{thm:cyclicwithoutcolluding}
    For the hierarchical secure aggregation with cyclic wrap-around association, when $n\leq K - 1$, the optimal rate region is given by 
    \[\mathcal{R}^*=\left\{R_X\geq\frac{1}{n},R_Y\geq\frac{1}{n},R_Z\geq\frac{1}{n},R_{Z_\Sigma}\geq\max\left\{1,\frac{K}{n}-1\right\}\right\}.\]
\end{thm}
Note that $R_{Z_\Sigma}\geq K/n-1$ is caused by the server security.
To ensure consistency with the notation used in this paper, it is crucial to clarify the definition of $R_X$ in relation to the results presented in \cite{2024zhangwansunwangoptimal}. We clarify the definition of $R_X$ in relation to the results from \cite{2024zhangwansunwangoptimal}. In that work, $R_X$ is defined as the total message length transmitted by each user, normalized by the length of $W_i$. Therefore, the result presented in \cite{2024zhangwansunwangoptimal} leads to the condition $R_X\geq1$. In our notation, $R_X$ reflects the message length over one user-to-relay link, normalized by the length of $W_i$.  

\subsection{Network function computation}
In this subsection, we present the basic model of NFC. We establish that the computation task in HSA, without any security constraints, is equivalent to an NFC problem. The HSA model can thus be viewed as an NFC model augmented with the additional security constraints.

In NFC, a \emph{network} is characterized by a directed acyclic graph $G=(\mathcal{V},\mathcal{E})$, where $\mathcal{V}$ is a finite vertex set and $\mathcal{E}$ is an edge set.
For every edge $e\in \mathcal{E}$, we use $\tail(e)$ and $\head(e)$ to denote the tail node and the head node of $e$.
For any vertex $v\in \mathcal{V}$, let $\In(v)=\set{e\in E;\head(e)=v}$ and $\Out(v)=\set{e\in E;\tail(e)=v}$, respectively.
A network $\cN$ over $G$ contains a set of \emph{source nodes} $S=\set{s_1,s_2,\ldots,s_N}\subseteq \cV$ and a \emph{sink node} $\gamma \in \cV\setminus S$. Such a network is denoted by $\cN=(G,S,\gamma)$. Without loss of
generality, we assume that every source node has no incoming
edges. We further assume that there exists a directed path from every node $u \in \cV\setminus \set{\gamma}$ to $\gamma$ in $G$. Then it follows from the
acyclicity of $G$ that the sink node $\gamma$ has no outgoing edges.
The sink node $\gamma$ needs to compute a \emph{target function} $f$ of the form \[f: \cA^N\longrightarrow\mathcal{O},\]
where $\cA$ and $\cO$ are finite alphabets, and the $i$-th argument of $f$ is generated at the source node $  s_i$. 

Let $k$ and $n$ be two positive integers, and let $\cB$ be a finite alphabet. A $(k,n)$ network code consists of a set of \emph{local encoding functions} $\{\theta_e:e\in\mathcal{E}\}$, where
\begin{equation}
  \theta_e:
  \begin{cases}
    \mathcal{A}^k\longrightarrow \mathcal{B}^n, & \mbox{if $\tail(e)=   s_i$ for some $i$}; \\
    \prod\limits_{d\in \In(\tail(e))}\mathcal{B}^n\longrightarrow\mathcal{B}^n, & \mbox{otherwise}.
  \end{cases}
\end{equation}
These functions  determine the messages transmitted by the edges recursively. Let $\bu_e \in \cB^n$ denote the message transmitted by the edge $e$. Then $\bu_e$ is a function of all source messages, i.e., for each $e\in \mathcal{E}$,
\begin{equation}
  \bu_e=g_e(\bx_S)=g_e(\bx_1,\cdots,\bx_N),
\end{equation}
where $g_e$ is the \emph{global encoding function} of edge $e$. For $E\subseteq \cE$, let
$g_E(\bx_S) = (g_e(\bx_S) : e\in E)$.
If $\tail(e)=  s_i\in S$, then
\begin{equation*}
    g_e(\bx_S) = \theta_e(\bx_i);
\end{equation*}
if $\tail(e)=u\in\mathcal{V}\backslash S$, then
\begin{equation*}
    g_e(\bx_S) = \theta_e(g_{\In(u)}(\bx_S)).
\end{equation*}
Moreover, the $(k,n)$ network code also includes a decoding function
\begin{equation*}
    \phi: \prod\limits_{e\in\In(\gamma)}\cB^n\longrightarrow\mathcal{O}^k
\end{equation*}
at the sink node $\gamma$. 
If $\phi(g_{\In(\gamma)}(\bx_S))=f(\bx_S)$ for all $\bx_S\in \cA^{Nk}$, we say that the network code
\emph{computes} $f$ and call  $k/n$ an \emph{achievable rate}.
The \emph{computing capacity} of the network $\mathcal{N}$ with the target function $f$ is defined as
\[C(\mathcal{N},f)\eqdef \sup \left\{\frac{k}{n}: \frac{k}{n} \textup{ is achievable}\right\}.
\]

\begin{figure}
    \centering
    \includegraphics[width=0.6\linewidth]{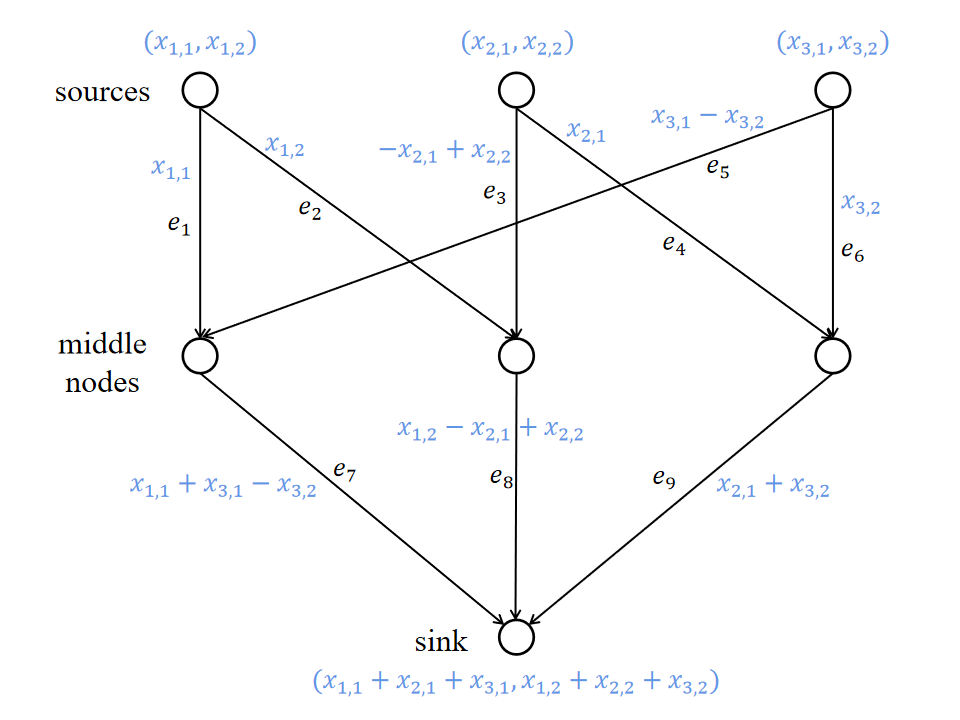}
    \caption{A $(2,1)$ network code over the network in Fig.~\ref{fig:HSA2NFC}}
    \label{fig:NFCsolution}
\end{figure}

We illustrate these definitions through the following example.
\begin{example}
    Consider the network shown in Fig.~\ref{fig:HSA2NFC}(b), which consists of $3$ source nodes, $3$ middle nodes, and $1$ sink node.  
    Let $\cA=\cB=\cO=\mathbb{F}_2$.
    The goal of the sink is to compute the sum of the three source messages, i.e., $f(x_1, x_2, x_3) = x_1 + x_2 + x_3$. A $(2,1)$ network code for this problem is illustrated in Fig.~\ref{fig:NFCsolution}. This code allows the target function to be computed twice while requiring only one message to be transmitted over each edge. The resulting computing rate is therefore $2/1 = 2$. Furthermore, this rate is optimal, as it achieves the cut-set bound for the given NFC problem. Hence, the computing capacity of the network in Fig.~\ref{fig:HSA2NFC}(b) is $2$.
\end{example}

In the HSA problem considered in this paper (illustrated in Fig.~\ref{fig:HSA2NFC}), the network topology can be directly mapped to a three-layer NFC model: users correspond to sources, relays to middle nodes, and the aggregation server to the sink. The target function in the corresponding NFC formulation is the sum of the source inputs.
The security requirements in HSA can be incorporated into the NFC framework as a form of \emph{node security}. Specifically, a $(k,n)$ network code is $(0,T_h,T_u)$-secure if any coalition of $T_h$ middle nodes and $T_u$ sources can learn no information about the inputs of the non-colluding sources. 
If a $(0, T_h, T_u)$-secure $(k,n)$ network code exists for the corresponding NFC problem, it immediately yields a valid scheme for the original HSA problem. In such a scheme, the communication loads are given by $R_X = R_Y = \frac{n}{k}$, since each edge carries a message of length $n$ while the source messages have length $k$. In particular, the communication load in the HSA setting is inversely proportional to the computing rate of the associated NFC code.

Next, we introduce several key notations and results in NFC.
Throughout the remainder of this paper, we assume that $\mathcal{A} = \mathcal{B} = \mathcal{O} = \mathbb{F}_q$.
An edge sequence $(e_1,e_2,\cdots,e_n)$ is called a \emph{path} from node $u$ to node $v$ if $\tail(e_1)=u,\head(e_n)=v$ and $\tail(e_{i+1})=\head(e_i)$ for $i=1,2,\cdots,n-1$.
For two nodes $u,v\in \mathcal{V}$, a \emph{cut} of them is an edge set $C$ such that every path from $u$ to $v$ contains at least one edge in $C$.
 If $C$ is a cut of $\gamma$ and some source node $\sigma_i$, then we simply call it a \emph{cut of the network}.
Let $\Lambda(\mathcal{N})$ be the collection of all cuts of the network $\mathcal{N}$.

A $(k,n)$ network code is \emph{linear} if every local encoding function $\theta_e$ is a linear function.
For the case where the target function is the sum over $\mathbb{F}_q$, Appuswamy \textit{et al.}~\cite{2011AFKZ} established the following theorem.
\begin{thm}[Theorem~III.2\cite{2011AFKZ}]\label{thm:NFC_cutset}
    For any network $\mathcal{N}$, if the target function is the sum over $\mathbb{F}_q$, then the computing capacity is given by
    \[C(\cN,f)=\mincut (\cN,f)\eqdef\min_{C\in\Lambda(\cN)}|C|.\]
\end{thm}

In \cite{2011AFKZ}, it was shown that this upper bound for capacity is achievable using a $(k,1)$ linear network code, where the source message has length $k$ and each edge transmits a message of length $1$. Since $n = 1$ and each local encoding function $\theta_e$ is linear, the message $u_e \in \mathbb{F}_q$ transmitted on an edge $e$ is a linear combination of the messages available at its tail node $\text{tail}(e)$. Specifically,
\begin{equation}\label{eq:linlocalenc}
  u_e=
  \begin{cases}
    \sum\limits_{j=1}^{k} x_{ij} p_{(i,j),e}, & \mbox{if $\tail(e)=   s_i$ for some $i$}; \\
    \sum\limits_{d\in \In(\tail(e))} u_d p_{d,e}, & \mbox{otherwise},
  \end{cases}
\end{equation}
where $p_{(i,j),e}$ and $p_{d,e}$ in $\mathbb{F}_{q}$ are referred to as the \emph{local encoding coefficients}. Here, $p_{(i,j),e} = 0$ if $e$ is not an outgoing edge of source node $s_i \in S$, and similarly, $p_{d,e} = 0$ if $e$ is not an outgoing edge of the node $\text{head}(d)$.

In a $(k,1)$ linear network code, the transmissions can be expressed by matrix representations.
For a subset of links $\rho \subseteq \mathcal{E}$, define the matrix $\mathbf{A}_\rho = (a_{d,e})_{d \in \rho, e \in \mathcal{E}}$, where
\begin{equation*}
  a_{d,e}=
  \begin{cases}
    1, & \mbox{if $d=e$}; \\
    0, & \mbox{otherwise}.
  \end{cases}
\end{equation*}
Let $\mathbf{P} \triangleq (p_{d,e})_{d \in \mathcal{E}, e \in \mathcal{E}}$ denote the \emph{local encoding coefficient matrix}. For each source node $s_i$, define $\textbf{P}_i \triangleq (p_{(i,j),e})_{j \in [k], e \in \mathcal{E}} = \textbf{E}_i^T \cdot \textbf{A}_{\Out(i)}$, where $\textbf{E}_i$ is the source encoding matrix at source node $s_i$. 

Using these matrices, the message received by the sink node can be expressed as:
\begin{equation*}
    \bx\cdot\begin{pmatrix}
        \textbf{E}_1^T\cdot \textbf{A}_{Out(1)}\\
        \textbf{E}_2^T\cdot \textbf{A}_{Out(2)}\\
        \vdots\\
        \textbf{E}_N^T\cdot \textbf{A}_{Out(N)}
    \end{pmatrix}\cdot (\textbf{I}-\textbf{P})^{-1}\cdot \textbf{A}_{In(\gamma)}^T.\footnote{The matrix representation is applicable to all directed acyclic networks. The corresponding transfer matrix, given by $(\textbf{I}-\textbf{P})^{-1}$, originates from the series $\textbf{I}+\textbf{P}+\textbf{P}^2+\cdots$, which cumulatively accounts for the transfer across all possible paths.}
\end{equation*}
Moreover, there is a linear decoding function $\phi :\prod_{\In(\gamma)} \F_q \to \F_q^{k}$ such that for all $\bx_S \in \F_q^{Nk}$,
\begin{equation*}
    \bx\cdot\begin{pmatrix}
        \textbf{E}_1^T\cdot \textbf{A}_{Out(1)}\\
        \textbf{E}_2^T\cdot \textbf{A}_{Out(2)}\\
        \vdots\\
        \textbf{E}_N^T\cdot \textbf{A}_{Out(N)}
    \end{pmatrix}\cdot (\textbf{I}-\textbf{P})^{-1}\cdot \textbf{A}_{In(\gamma)}^T\cdot \textbf{D}^T=\sum_{i=1}^s\bx_i,
\end{equation*}
where $\textbf{I}\in\mathbb{F}_q^{|\cE|\times|\cE|}$ is an identity matrix.
In such a case, we say that the linear network code enables the sink node to compute the target function with rate $k$.

We illustrate these definitions using the following example.
\begin{example}
    The $(2,1)$ network code shown in Fig.~\ref{fig:NFCsolution} can be described in terms of three types of matrices. First, the encoding matrices at the source nodes $s_i$, for $i = 1, 2, 3$, are given by:
    \begin{equation*}
        \textbf{E}_1=\begin{pmatrix}
            1&0\\
            0&1
        \end{pmatrix},\ \ 
        \textbf{E}_2=\begin{pmatrix}
            -1&1\\
            1&1
        \end{pmatrix},\ \ 
        \textbf{E}_3=\begin{pmatrix}
            1&-1\\
            0&1
        \end{pmatrix}.
    \end{equation*}
    The matrices $\textbf{A}_\rho$ are determined by the network topology. For the network in Fig.~\ref{fig:NFCsolution}, we have
    \begin{align*}
        &\textbf{A}_{\Out(1)}=\begin{pmatrix}
            1&0&0&0&0&0&0&0&0\\
            0&1&0&0&0&0&0&0&0
        \end{pmatrix},\\
        &\textbf{A}_{\Out(2)}=\begin{pmatrix}
            0&0&1&0&0&0&0&0&0\\
            0&0&0&1&0&0&0&0&0
        \end{pmatrix},\\
        &\textbf{A}_{\Out(3)}=\begin{pmatrix}
            0&0&0&0&1&0&0&0&0\\
            0&0&0&0&0&1&0&0&0
        \end{pmatrix},\\
        &\textbf{A}_{\In(\gamma)}=\begin{pmatrix}
            0&0&0&0&0&0&1&0&0\\
            0&0&0&0&0&0&0&1&0\\
            0&0&0&0&0&0&0&0&1
        \end{pmatrix}.\\
    \end{align*}
    Since each middle node simply sums the messages it receives, the local encoding coefficient $p_{d,e}$ equals $1$ if and only if $\mathrm{head}(d) = \mathrm{tail}(e)$. Specifically,
    \begin{equation*}
        p_{e_1,e_7}=p_{e_2,e_8}=p_{e_3,e_8}=p_{e_4,e_9}=p_{e_5,e_7}=p_{e_6,e_9}=1,
    \end{equation*}
    and else $p_{e_i,e_j}=0$.
    For $\bx=(x_{1,1},x_{1,2},x_{2,1},x_{2,2},x_{3,1},x_{3,2})$, the message received by the sink node is 
    \begin{subequations}
        \begin{align*}
            &\bx\cdot\begin{pmatrix}\textbf{E}_1^T\cdot \textbf{A}_{Out(1)}\\\textbf{E}_2^T\cdot \textbf{A}_{Out(2)}\\\textbf{E}_3^T\cdot \textbf{A}_{Out(3)}\end{pmatrix}\cdot (\textbf{I}-\textbf{P})^{-1}\cdot \textbf{A}_{In(\gamma)}^T
            =\bx\cdot\begin{pmatrix}
                1&0&0&0&1&-1\\
                0&1&-1&1&0&0\\
                0&0&1&0&0&1
            \end{pmatrix}^T\\
            =&\begin{pmatrix}
                x_{1,1}+x_{3,1}-x_{3,2}&
                x_{1,2}-x_{2,1}+x_{2,2}&
                x_{2,1}+x_{3,2}
            \end{pmatrix}.
        \end{align*}
    \end{subequations}
    Finally, the decoding matrix $\textbf{D}$ is given by:
    \begin{equation*}
        \textbf{D}=\begin{pmatrix}
            1&0\\
            0&1\\
            1&1
        \end{pmatrix}.
    \end{equation*}
\end{example}

\section{Main results}\label{sec:mainresults}
In this section, we present our main results on the rate region for hierarchical secure aggregation under $(0,T_h,T_u)$-security.
We begin by establishing fundamental lower bounds on the communication loads, $R_X$ and $R_Y$, which are derived from the cut-set bound. The detail proof of Theorem~\ref{thm:lowerboundcommuni} is presented in Section~\ref{sec:converse}-A.
\begin{thm}\label{thm:lowerboundcommuni}
    For the hierarchical secure aggregation problem $(N,K,n,T_u,T_h,0)$, where $N \ge K$ and $K$ divides $Nn$, the communication loads, $R_X$ and $R_Y$, must satisfy the following inequalities:
    \[R_X \ge \frac{1}{n}, \quad R_Y \ge \frac{1}{n}.\]
\end{thm}

For $0 < T_h \le K-n$, define $n(\mathcal{N}_{N,K,n},T_h)$ as the minimum number of users such that there exist $K-T_h-n+1$ relays that connect only to these users. Specifically,
\[n(\mathcal{N}_{N,K,n},T_h) = \min_{\mathcal{H} \subseteq [K]: |\mathcal{H}| = K-T_h-n+1} \left| \bigcup_{j \in \mathcal{H}} \mathcal{U}_j \right|,\]
where $\cU_j$ is the set of users associated with the relay $j$. 

\begin{thm}\label{thm:unachievable}
    For the hierarchical secure aggregation problem $(N,K,n,T_u,T_h,0)$, if either $T_h \ge K-n+1$ or $0 < T_h \le K-n,T_u \ge n(\mathcal{N}_{N,K,n}, T_h)$, there exists no $(0,T_h,T_u)$-secure scheme that achieves the communication load pair $(R_X,R_Y)=(1/n,1/n)$.
\end{thm}
The proof is presented in Section~\ref{sec:converse}-B.
Although the lower bound derived for communication load is intuitively evident, it can be achieved for any user-relay connection if $T_h \le K-n$ and $T_u < n(\mathcal{N}_{N,K,n}, T_h)$. The achievable scheme is presented in Section~\ref{sec:scheme}-A and constructed by modifying an $(n,1)$ network code for the corresponding NFC problem.

\begin{thm}\label{thm:scheme(1,u-1)}
    For the hierarchical secure aggregation problem $(N,K,n,T_u,T_h,0)$, if $0 < T_h \le K-n$ and $T_u \le n(\mathcal{N}_{N,K,n}, T_h)-1$, then there exists a $(0,T_h,T_u)$-secure scheme that achieves the communication load pair $(R_X,R_Y)=(1/n,1/n)$ with $(R_Z, R_{Z_{\Sigma}}) = (1, N-1)$.
\end{thm}

Theorem~\ref{thm:scheme(1,u-1)} demonstrates that the optimal communication load is achievable as long as $T_h \le K-n$ and $T_u \le n(\mathcal{N}_{N,K,n}, T_h)-1$, when the key rates are not of great concern. Conversely, Theorem~\ref{thm:unachievable} implies that exceeding either threshold makes the optimal communication load unattainable. Therefore, a key objective becomes minimizing the key rates $R_Z$ and $R_{Z_{\Sigma}}$ while maintaining optimal communication loads. We thus derive the following lower bound on the key rates. The detail proof of Theorem~\ref{thm:lowerboundR_Z} is presented in Section~\ref{sec:converse}-C.

\begin{thm}\label{thm:lowerboundR_Z}
    For the hierarchical secure aggregation problem $(N,K,n,T_u,T_h,0)$, let $0 < T_h \le K-n$ and $T_u \le n(\mathcal{N}_{N,K,n}, T_h)-1$. If there is a $(0,T_h,T_u)$-secure scheme that achieves the communication load pair $(R_X,R_Y)=(1/n,1/n)$, then we must have
    \[R_Z \ge \min\left\{\frac{T_h}{n},1\right\}.\]
    Furthermore, if $T_h m + T_u < N$, then \[R_{Z_{\Sigma}} \ge \min\left\{\frac{T_h(T_u+m)}{n},\frac{T_un+T_hm}{n}\right\}.\]
\end{thm}

When $T_h=1$, the lower bound above can be achieved for a special class of networks. 
In particular, in a \emph{multiple cyclic network}, for each user $i\in[N]$, $\cH_i=\{\Mod(i,K),\Mod(i-1,K),\cdots,\Mod(i-n+1,K)\}.$
For each relay $j\in[K]$, 
$\cU_j=\cup_{p\in[0:\frac{N}{K}-1]}\{\Mod(j,K)+pK,\Mod(j+1,K)+pK,\cdots,\Mod(j+n-1,K)+pK\}.$
Here, we always assume $N/K$ is an integer. If $N=K$, we say the network is a \emph{cyclic network}. When $T_h=1,T_u+m<N$, Theorem~\ref{thm:lowerboundR_Z}
 implies $R_Z\geq1/n$ and $R_{Z_\Sigma}\geq (T_u+m)/n$.

\begin{thm}\label{thm:scheme(1/n)}
    For the hierarchical secure aggregation problem $(N,K,n,T_u,T_h,0)$, let $T_h=1,T_u\leq n(\cN_{N,K,n},1)-1$, such that $T_u+m\leq \min\{N-1,K-n\}$. For a multiple cyclic network, there exists a $(0,1,T_u)$-secure scheme that achieves the communication load pair $(R_X,R_Y)=(1/n,1/n)$ with $(R_Z,R_{Z_{\Sigma}})=(\frac{1}{n},\frac{T_u+m}{n})$.
\end{thm}

\emph{Analysis of Results:} We begin by comparing our results with those presented in \cite{2024zhangwansunwangoptimal} and \cite{2025HSAcyclic}. For a tree network topology, as analyzed in Theorem~\ref{thm:lowerboundcommuni} and Theorem~\ref{thm:lowerboundR_Z}, when $T$ colluding users are present (i.e. (0,1,T)-secure), we obtain a lower bound of $R_X \geq 1$, $R_Y \geq 1$, $R_Z \geq 1$, and $R_{Z_\Sigma} \geq m + T$ for cases where $T < (K-1)m$. Furthermore, our results indicate that no secure HSA scheme exists when $T \ge (K-1)m$. This directly aligns with the results established in \cite{2024zhangwansunwangoptimal}, as formally summarized in Theorem~\ref{thm:treewithcolluding}. Note that, consistent with our focus on relay security, where the server is not involved in collusion, the term $R_{Z_\Sigma} \ge U + T - 1$ is not considered here.

Next, we examine cyclic networks without colluding users ($T_h=1,T_u=0$). Based on the aforementioned theorems, we obtain the following lower bounds: $R_X \ge 1/n$, $R_Y \ge 1/n$, $R_Z \ge 1/n$, and $R_{Z_\Sigma} \ge m/n = 1$. These results are consistent with the findings presented in \cite{2025HSAcyclic}, as formalized in Theorem~\ref{thm:cyclicwithoutcolluding}. Similarly, the term $R_{Z_\Sigma} \ge K/n - 1$ is omitted, as we do not address server security in this paper.

Finally, we analyze cyclic networks in the presence of $T$ colluding users. Specifically, we consider the scenario where $N = K$, $n = m < K$, and $T_h = 1$, $T_u = T$. Our results show that for $T \ge N-1$, there are no $(0, 1, T)$-secure HSA schemes. Moreover, when $T \le N-2$, we have the lower bounds $R_X \ge 1/n$, $R_Y \ge 1/n$, $R_Z \ge 1/n$, and $R_{Z_\Sigma} \ge (T + m)/n$. Furthermore, Theorem~\ref{thm:scheme(1,u-1)} and Theorem~\ref{thm:scheme(1/n)} demonstrate the existence of a $(0, 1, T)$-secure HSA scheme, achieving the lower bound, for the specific case of $T \le N-2n$.

In Section~\ref{sec:example}, we provided a specific example with $N=K=3$, $n=m=2$, and $T_h=T_u=1$, demonstrating the tightness of the lower bounds $R_Z \ge 1$ and $R_{Z_\Sigma} \ge 2$. Building upon this example, the following theorem generalizes this finding to any cyclic network with $n=m=2$.

\begin{thm}\label{thm:casen=m=2}
    Consider the hierarchical secure aggregation problem $(N,K,n,T_u,T_h,0)$ defined over a cyclic network, with parameters $N=K$, $n=2$, $T_h=1$, and $T_u=N-2$. If there exists a $(0,1,N-2)$-secure HSA scheme that achieves the communication load pair $(R_X,R_Y)=(1/n,1/n)$, then it must hold:
    \[R_Z\geq 1 \quad \text{and} \quad R_{Z_\Sigma}\geq N-1.\]
\end{thm}

For a cyclic network with $n=m=2$, our previous analysis indicates that if $T \leq N-4$, the optimal rate region is characterized by $R_X \ge 1/2$, $R_Y \ge 1/2$, $R_Z \ge 1/2$, and $R_{Z_\Sigma} \ge T/2+1$. When $T=N-2$, the optimal rate region shifts to $R_X \ge 1/2$, $R_Y \ge 1/2$, $R_Z \ge 1$, and $R_{Z_\Sigma} \ge N-1$. However, for the case where $T=N-3$, while we have derived the lower bounds $R_Z \ge 1/2$ and $R_{Z_\Sigma} \ge T/2+1$, a matching scheme has not yet been presented. Therefore, we introduce a novel scheme to demonstrate the tightness of these lower bounds.

\begin{thm}\label{thm:caseT=N-3}
    For the hierarchical secure aggregation problem $(N,K,n,T_u,T_h,0)$ over a cyclic network, with parameters $N=K$, $n=2$, $T_h=1$, and $T_u=N-3$, there exists a $(0,1,N-3)$-secure HSA scheme that attains the optimal communication load with key rates $R_Z=1/2$ and $R_{Z_\Sigma}=(T_u+m)/2=(N-1)/2$.
\end{thm}

Collecting the results above, we have comprehensively addressed the HSA problem with $(0,1,T)$-security for a cyclic network where $n=m=2$. This leads to the following corollary:

\begin{cor}
    For the HSA problem over a cyclic network with $n=m=2$ and $(0,1,T)$-security:
    \begin{itemize}
        \item If $T \leq N-3$, the optimal key rates are $R_Z=\frac{1}{2}$ and $R_{Z_\Sigma}=\frac{T}{2}+1$.
        \item If $T=N-2$, the optimal key rates are $R_Z=1$ and $R_{Z_\Sigma}=N-1$.
        \item If $T \geq N-2$, a $(0,1,T)$-secure HSA scheme that attains the optimal communication load does not exist.
    \end{itemize}
\end{cor}

\section{Converse proof}\label{sec:converse}
In this section, we establish converse results using information-theoretic techniques. We begin by presenting a lemma, instrumental in proving lower bounds on the communication loads $R_X$ and $R_Y$. 

\begin{lem}[Lemma~1 in \cite{2025HSAcyclic}]\label{lem:xijyj}
    For every user $i\in[N]$, it holds that
    \begin{align}
        &H(\{X_{i,j}\}_{j\in\cH_i}|\{W_{i'},Z_{i'}\}_{i'\in[N]\backslash\{i\}})\geq L,\\
        &H(\{Y_{j}\}_{j\in\cH_i}|\{W_{i'},Z_{i'}\}_{i'\in[N]\backslash\{i\}})\geq L.
    \end{align}
\end{lem}
Lemma~\ref{lem:xijyj} has been proven in \cite{2025HSAcyclic} for the cyclic network. Since the proof is only based on the cut-set bound, it is applicable to any user-relay association.

\subsection{Proof of Theorem~\ref{thm:lowerboundcommuni}}
\begin{itemize}
    \item \emph{Proof of} $R_X\geq\frac{1}{n}$: For every $i\in[N]$, we have
    \begin{subequations}
        \begin{align}
            nR_XL&\geq nH(X_{i,j})\geq H(\{X_{i,j}\}_{j\in\cH_i})\\
            &\geq H(\{X_{i,j}\}_{j\in\cH_i}|\{W_{i'},Z_{i'}\}_{i'\in[N]\backslash\{i\}})\geq L,
        \end{align}
    \end{subequations}
    which implies that $R_X\geq \frac{1}{n}$.
    \item \emph{Proof of} $R_Y\geq\frac{1}{n}$: Similarly to the above case, for every $i\in[N]$, we have
    \begin{subequations}
        \begin{align}
            nR_YL&\geq nH(Y_j)\geq H(\{Y_{j}\}_{j\in\cH_i})\\
            &\geq H(\{Y_j\}_{j\in\cH_i}|\{W_{i'},Z_{i'}\}_{i'\in[N]\backslash\{i\}})\geq L,
        \end{align}
    \end{subequations}
    which implies that $R_Y\geq \frac{1}{n}$.
\end{itemize}

Note that the lower bounds derived above do not require the secure conditions. 

\subsection{Proof of Theorem~\ref{thm:unachievable}}
The proof can be divided into two cases, 1) $T_h\geq K-n+1$ and 2) $0<T_h\leq K-n, T_u\geq n(\cN_{N,K,n}, T_h)$.

For the first case, if $T_h\geq K-n+1$, let $\cT_h\subseteq[K]$, where $|\cT_h|=T_h$, be the set of colluding relays. For any colluding users $\cT_u\subseteq[N]$, where $|\cT_u|=T_u$, from the definition of $(0,T_h,T_u)$-security, we have

\begin{subequations}\label{eq:unachi_case1}
    \begin{align}
         0&=I(W_{[N]};\{X_{i,j}:i\in\cU_j\}_{j\in\cT_h}|\{W_i,Z_i\}_{i\in\cT_u})\\
         &=I\left(W_{[N]},\sum_{i=1}^{N}W_i;\{X_{i,j}:i\in\cU_j\}_{j\in\cT_h},\{Y_j\}_{j\in\cT_h}|\{W_i,Z_i\}_{i\in\cT_u}\right)\\
         &\geq I\left(\sum_{i=1}^{N}W_i;\{Y_j\}_{j\in\cT_h}|\{W_i,Z_i\}_{i\in\cT_u}\right)\\
         &=I\left(\sum_{i=1}^{N}W_i;\{Y_j\}_{j\in\cT_h},\{W_i,Z_i\}_{i\in\cT_u}\right)-I\left(\sum_{i=1}^{N}W_i;\{W_i,Z_i\}_{i\in\cT_u}\right)\\
         &=I\left(\sum_{i=1}^{N}W_i;\{Y_j\}_{j\in\cT_h},\{W_i,Z_i\}_{i\in\cT_u}\right)\\
         &\geq I\left(\sum_{i=1}^{N}W_i;\{Y_j\}_{j\in\cT_h}\right)\\
         &=I\left(\sum_{i=1}^{N}W_i;\{Y_j\}_{j\in[K]}\right)-I\left(\sum_{i=1}^{N}W_i;\{Y_j\}_{j\in[K]\backslash\cT_h}|\{Y_j\}_{j\in\cT_h}\right)\\
         &\geq I\left(\sum_{i=1}^{N}W_i;\{Y_j\}_{j\in[K]}\right)-H(\{Y_j\}_{j\in[K]\backslash\cT_h})\\
         &\geq L-(K-T_h)\frac{L}{n}>0,
    \end{align}
\end{subequations}
leading to a contradiction. Equation (\ref{eq:unachi_case1}a) follows from the $(0, T_h,T_u)$-security constraint.  The second equality in (\ref{eq:unachi_case1}b) is a consequence of the fact that $\sum_{i=1}^{N}W_i$ and $Y_j$ are functions of $W_{[N]}$ and $\{X_{i,j}\}_{i\in\cU_j}$, respectively. In (\ref{eq:unachi_case1}d), the second term vanishes due to the independence between $W_i$ (which are uniformly distributed over a finite field) and $Z_i$. Finally, in (\ref{eq:unachi_case1}h), the first term equals $L$ because the server is able to decode the sum from $\{Y_j\}_{j\in[K]}$.  Furthermore, due to the communication load optimization, each $Y_j$ has a length of $\frac{L}{n}$, implying that the second term is upper bounded by $(K-T_h)\frac{L}{n}$.

For the second case, where $0<T_h\leq K-n$ and $T_u\geq n(\cN_{N,K,n}, T_h)$, let $\cT_h$ denote an arbitrary colluding relay set of size $T_h$.  Furthermore, for any subset $\cH \subseteq [K]\backslash\cT_h$ of size $K-T_h-n+1$, we define $\cT_u = \bigcup_{j\in\cH}\cU_j$.  By the definition of $n(\cN_{N,K,n}, T_h)$, we have $|\cT_u|\geq n(\cN_{N,K,n}, T_h)$.  With this specific choice of $\cT_h$ and $\cT_u$, we establish the following chain of inequalities:
\begin{subequations}\label{eq:unachi_case2}
    \begin{align}
        0&=I(W_{[N]};\{X_{i,j}:i\in\cU_j\}_{j\in\cT_h}|\{W_i,Z_i\}_{i\in\cT_u})\\
        &= I\left(W_{[N]},\sum_{i=1}^{N}W_i;\{X_{i,j}:i\in\cU_j\}_{j\in\cT_h},\{Y_j\}_{j\in\cT_h}|\{W_i,Z_i\}_{i\in\cT_u},\{Y_{j}\}_{j\in\cH}\right)\\
        &\geq I\left(\sum_{i=1}^{N}W_i;\{Y_j\}_{j\in\cT_h}|\{W_i,Z_i\}_{i\in\cT_u},\{Y_{j}\}_{j\in\cH}\right)\\
        &= I\left(\sum_{i=1}^{N}W_i;\{Y_j\}_{j\in\cT_h},\{W_i,Z_i\}_{i\in\cT_u},\{Y_{j}\}_{j\in\cH}\right)-I\left(\sum_{i=1}^{N}W_i;\{W_i,Z_i\}_{i\in\cT_u},\{Y_{j}\}_{j\in\cH}\right)\\
        &=I\left(\sum_{i=1}^{N}W_i;\{Y_j\}_{j\in\cT_h},\{W_i,Z_i\}_{i\in\cT_u},\{Y_{j}\}_{j\in\cH}\right)\\
        &\geq I\left(\sum_{i=1}^{N}W_i;\{Y_j\}_{j\in\cT_h},\{Y_{j}\}_{j\in\cH}\right)\\
        &=I\left(\sum_{i=1}^{N}W_i;,Y_{[H]}\right)-I\left(\sum_{i=1}^{N}W_i;\{Y_{j}\}_{j\in[K]\backslash(\cH\cup\cT_h)}|\{Y_j\}_{j\in\cT_h},\{Y_{j}\}_{j\in\cH}\right)\\
        &\geq I\left(\sum_{i=1}^{U}W_i;Y_{[H]})-H(\{Y_{j}\}_{j\in[K]\backslash(\cH\cup\cT_h)}\right)
        \geq L-(n-1)\frac{L}{n}>0,
    \end{align}
\end{subequations}
which again leads to a contradiction. Equation (\ref{eq:unachi_case2}a) is a direct consequence of the $(0, T_h,T_u)$-security requirement. The second equality in (\ref{eq:unachi_case2}b) follows from the fact that $\sum_{i=1}^{N}W_i$ and $Y_j$ are functions of $W_{[N]}$ and $\{X_{i,j}\}_{j\in\cT_h,i\in\cU_j}$, respectively. Note that in (\ref{eq:unachi_case2}d), the second term equals to zero due to the independence of $W_i$ (uniformly distributed over a finite field) and $Z_i$. Finally, in (\ref{eq:unachi_case2}h), the first term equals $L$ because the server can successfully decode the sum from $\{Y_j\}_{j\in[K]}$. Moreover, since the communication load is optimized, each $Y_j$ has length $\frac{L}{n}$, implying that the second term is no larger than $(K-T_h-|\cH|)\frac{L}{n}=(K-T_h-(K-T_h-n+1))\frac{L}{n}=(n-1)\frac{L}{n}$.

In conclusion, the preceding analysis demonstrates that, under the condition of optimal communication load, the $(0,T_h,T_u)$-security requirement and decodability cannot be simultaneously satisfied for both cases.

\subsection{Proof of Theorem~\ref{thm:lowerboundR_Z}}
In the following, we first introduce several definitions that are instrumental in the converse proof. Let $W_{i,j}$ and $Z_{i,j}$ represent the symbols in $W_i$ and $Z_i$ which appear in the encoding of $X_{i,j}$, respectively. Consequently, the following relationships hold:
\begin{align}
    &H(X_{i,j}|W_{i,j},Z_{i,j})=0,\label{eq:defxij}\\
    &H(Z_{i,j})=H(X_{i,j}|W_{i,j}),\label{eq:defzij}\\
    &H(W_{i,j})=H(X_{i,j}|Z_{i,j}).\label{eq:defwij}
\end{align}
Furthermore, we establish the following properties:

\begin{prop}\label{prop:zijwij}
    Given $X_{i,j}$ as the message transmitted from user $i$ to relay $j$, and with $Z_{i,j}$ and $W_{i,j}$ defined as above, the following equations are satisfied:
    \begin{align}
        &H(Z_{i,j}|W_{i,j},X_{i,j})=0,\label{eq:propzij}\\
        &H(W_{i,j}|Z_{i,j},X_{i,j})=0.\label{eq:propwij}
    \end{align}
\end{prop}
\begin{pf}
    We first prove (\ref{eq:propzij}). From the definition of $Z_{i,j}$, we have
    \begin{subequations}\label{eq:pfpropzij}
        \begin{align}
            H(Z_{i,j}) &\geq H(Z_{i,j}|W_{i,j}) \\
            &= H(Z_{i,j},W_{i,j}) - H(W_{i,j}) \\
            &= H(Z_{i,j},W_{i,j},X_{i,j}) - H(W_{i,j}) \\
            &= H(Z_{i,j},W_{i,j},X_{i,j}) - H(W_{i,j},X_{i,j}) + H(W_{i,j},X_{i,j}) - H(W_{i,j}) \\
            &= H(Z_{i,j}|X_{i,j},W_{i,j}) + H(X_{i,j}|W_{i,j}) \\
            &= H(Z_{i,j}|X_{i,j},W_{i,j}) + H(Z_{i,j}),
        \end{align}
    \end{subequations}
    where equation (\ref{eq:pfpropzij}c) follows from (\ref{eq:defxij}), which reflects that $X_{i,j}$ is a function of $W_{i,j}$ and $Z_{i,j}$. Equation (\ref{eq:pfpropzij}f) follows directly from (\ref{eq:defzij}). Consequently, we obtain (\ref{eq:propzij}). Similarly, we can prove (\ref{eq:propwij}) as follows:
    \begin{subequations}
        \begin{align}
            H(W_{i,j}) &\geq H(W_{i,j}|Z_{i,j}) \\
            &= H(Z_{i,j},W_{i,j}) - H(Z_{i,j}) \\
            &= H(Z_{i,j},W_{i,j},X_{i,j}) - H(Z_{i,j}) \\
            &= H(Z_{i,j},W_{i,j},X_{i,j}) - H(Z_{i,j},X_{i,j}) + H(Z_{i,j},X_{i,j}) - H(Z_{i,j}) \\
            &= H(W_{i,j}|X_{i,j},Z_{i,j}) + H(X_{i,j}|Z_{i,j}) \\
            &= H(W_{i,j}|X_{i,j},Z_{i,j}) + H(W_{i,j}),
        \end{align}
    \end{subequations}
    which implies (\ref{eq:propwij}).
\end{pf}
Now we are ready to derive the lower bound for $R_Z$ when the communication load is optimized.

\emph{Proof of} $R_Z \ge \min\{T_h/n,1\}$: 
If $T_h\geq n$, then $R_Z\geq1$. Otherwise, we have $R_Z\geq T_h/n$. 

From the security constraint, for any subset of relays $\cT_h$ and any subset of users $\cT_u$ with $|\cT_h|\leq T_h,|\cT_u|\leq T_u$ and $0<T_h\leq K-n,T_u\leq n(\cN_{N,K,n})-1$, we have
\begin{equation*}
    I(\{X_{i,j}:i\in\cU_j\}_{j\in\cT_h};\{W_{i,j}:i\in\cU_j\}_{j\in\cT_h}|\{W_i,Z_i\}_{i\in\cT_u})=0.
\end{equation*}
When $T_h<n$, without loss of generality, we assume  $\cT_h\subseteq\cH_1$. Then, we must have
\[I(\{X_{1,j}\}_{j\in\cT_h};\{W_{1,j}\}_{j\in\cT_h}|\{W_i,Z_i\}_{i\in\cT_u})=0.\]
Note that
\begin{subequations}\label{eq:lowerbound_Rz}
    \begin{align}
        &I(\{X_{1,j}\}_{j\in\cT_h};\{W_{1,j}\}_{j\in\cT_h}|\{W_i,Z_i\}_{i\in\cT_u})\\=&H(\{X_{1,j}\}_{j\in\cT_h}|\{W_i,Z_i\}_{i\in\cT_u})-H(\{X_{1,j}\}_{j\in\cT_h}|\{W_i,Z_i\}_{i\in\cT_u},\{W_{1,j}\}_{j\in\cT_h})\\
        \geq& H(\{X_{1,j}\}_{j\in\cT_h}|\{W_i,Z_i\}_{i\in\cT_u},Z_1)-H(\{X_{1,j}\}_{j\in\cT_h}|\{W_i,Z_i\}_{i\in\cT_u},\{W_{1,j}\}_{j\in\cT_h})\\
        =&H(\{X_{1,j},Z_{1,j}\}_{j\in\cT_h},|\{W_i,Z_i\}_{i\in\cT_u},Z_1)-H(\{X_{1,j},W_{1,j}\}_{j\in\cT_h}|\{W_i,Z_i\}_{i\in\cT_u},\{W_{1,j}\}_{j\in\cT_h})\\
        =&H(\{X_{1,j},Z_{1,j},W_{1,j}\}_{j\in\cT_h},|\{W_i,Z_i\}_{i\in\cT_u},Z_1)-H(\{X_{1,j},W_{1,j},Z_{1,j}\}_{j\in\cT_h}|\{W_i,Z_i\}_{i\in\cT_u},\{W_{1,j}\}_{j\in\cT_h})\\
        =&H(\{W_{1,j},Z_{1,j}\}_{j\in\cT_h}|\{W_i,Z_i\}_{i\in\cT_u},Z_1)-H(\{Z_{1,j},W_{1,j}\}_{j\in\cT_h}|\{W_i,Z_i\}_{i\in\cT_u},\{W_{1,j}\}_{j\in\cT_h})\\
        =&H(\{W_{1,j}\}_{j\in\cT_h},|\{W_i,Z_i\}_{i\in\cT_u},Z_1)-H(\{Z_{1,j}\}_{j\in\cT_h}|\{W_i,Z_i\}_{i\in\cT_u},\{W_{1,j}\}_{j\in\cT_h})\\
        =&H(\{W_{1,j}\}_{j\in\cT_h})-H(\{Z_{1,j}\}_{j\in\cT_h}|\{Z_i\}_{i\in\cT_u}),
    \end{align}
\end{subequations}
where the equality in (\ref{eq:lowerbound_Rz}d) holds because $\{Z_{1,j}\}_{j\in\cT_h}$ is determined by $Z_1$, (\ref{eq:lowerbound_Rz}e) is based on the properties of $W_{i,j},Z_{i,j}$ and $X_{i,j}$, which is presented in (\ref{eq:propzij}) and (\ref{eq:propwij}), the equality in (\ref{eq:lowerbound_Rz}f) is due to (\ref{eq:defxij}), the last equality in (\ref{eq:lowerbound_Rz}g) holds because of the independence of $W_i$ and $Z_i$. From (\ref{eq:lowerbound_Rz}), we directly obtain that 
\begin{align*}
    H(Z_1)\geq H(\{Z_{1,j}\}_{j\in\cT_h})
    \geq H(\{Z_{1,j}\}_{j\in\cT_h}|\{Z_i\}_{i\in\cT_u})
    \geq H(\{W_{1,j}\}_{j\in\cT_h}).
\end{align*}
Leveraging the definitions and properties of $W_{i,j}$, $X_{i,j}$, and $Z_{i,j}$, we can demonstrate that, given a scheme with $R_X=R_Z=\frac{1}{n}$, the following holds:
\begin{equation}\label{eq:sizewij}
    H(W_{i,j})=H(X_{i,j})=H(Z_{i,j})=\frac{L}{n}.
\end{equation}
Specifically, since $R_X=\frac{1}{n}$, implying $H(X_{i,j})=\frac{L}{n}$, a straightforward application of the cut-set bound yields:
\begin{equation*}
    \sum_{j\in\mathcal{H}_i}H(W_{i,j})\geq H(\{W_{i,j}\}_{j\in\mathcal{H}_i})\geq H(W_{i})=L. 
\end{equation*}
Combining this with the inequality $H(W_{i,j})=H(X_{i,j}|Z_{i,j})\leq H(X_{i,j})=\frac{L}{n}$,  we can obtain $H(W_{i,j})=L/n$ and $\{W_{i,j}\}_{j\in\cH_i}$ are mutually independent. This leads to
\[H(Z_1)\geq H(\{W_{1,j}\}_{j\in\cT_h})=\frac{T_hL}{n},\]
which implies $R_Z\geq\frac{T_h}{n}$.

When $T_h\geq n$, we can select $\cT_h=\cH_1$. By the same process as above, we obtain
\[H(Z_1)\geq H(\{W_{1,j}\}_{j\in\cH_1})=H(W_1)=L,\]
which implies $R_Z\geq 1$.

In the following, we will derive the lower bound for $R_{Z_\Sigma}$. We first present a lower bound for $H(\{Z_i\}_{i\in\cT_u})$, which implies that for a given $T_h$, the lower bound for $H(\{Z_i\}_{i\in\cT_u})$ is determined by the size of $\cT_u$.

\begin{lem}\label{lem:sizeofZTu}
    When the communication load is optimized, for any collusion user set $\mathcal{T}_u \subseteq [N]$ with $|\mathcal{T}_u| = T_u$, we have
    \[H(\{Z_i\}_{i\in\cT_u})\geq T_u\cdot\min\{L,T_hL/n\}.\]
\end{lem}

\begin{pf}
    If $T_h<n$, for any $\cS\subseteq[N]$ with $|\cS|=T_u-1$ and $i' \notin \cS$, we select $\cT_h$ such that $|\cT_h|=T_h$ and $\cT_h\subseteq\cH_{i'}$. Then, by an argument similar to (\ref{eq:lowerbound_Rz}), we can derive \[H(Z_{i'}|\{Z_i\}_{i\in\cS})\geq H(\{W_{i',j}\}_{j\in\cT_h})=\frac{T_hL}{n}.\] 
    If $T_h\geq n$, for any $\mathcal{T}_u$ and $i' \notin \mathcal{T}_u$, we select $\cT_h$ such that $\cT_h=\cH_{i'}$. Still by a similar argument to (\ref{eq:lowerbound_Rz}), we have
    \[H(Z_{i'}|\{Z_i\}_{i\in\cS})\geq H(W_{i'})= L.\] 
    Consequently, we can deduce that
    \begin{subequations}
        \begin{align}
            H(\{Z_{i}\}_{i\in\cT_u})\geq&\sum_{i'\in\cT_u}H(Z_{i'}|\{Z_{i}\}_{i\in\cT_u\backslash\{i'\}})\\
            =&|\cT_u|\cdot\frac{\min\{T_h,n\}L}{n},
        \end{align}
    \end{subequations}
    which completes the proof.
\end{pf}

The following lemma provides a lower bound on the entropy of the random keys, $\{Z_{i,j}\}$, contained in the messages received by colluding relays, conditioned on the random keys owned by the colluding users. Specifically, the lemma lower bounds the conditional entropy of $\{Z_{i,j}\}$ by a sum of conditional entropies of $\{X_{i,j}\}$.
\begin{lem}\label{lem:ZijtoXij}
    For any relay subset $\mathcal{T}_h \subseteq [K]$ and colluding user set $\mathcal{T}_u$, the following inequality holds: 
    \begin{equation}
        H(\{Z_{i,j}:i\in\cU_j\}_{j\in\cT_h}|\{Z_{i'}\}_{i'\in\cT_u})\geq\sum_{(i,j):j\in\cT_h,\ i\in\cU_j\backslash\cT_u}H(X_{i,j}|\{W_{i'},Z_{i'}\}_{i'\in[N]\backslash\{i\}}).
    \end{equation}
\end{lem}
\begin{pf}
    Recall that $\mathcal{U}_j$ denotes the set of users associated with relay $j$. For any $j \in [H]$, we have
    \begin{subequations}\label{eq:Zij}
        \begin{align}
            & H(\{Z_{i,j}:i\in\cU_j\}_{j\in\cT_h}|\{Z_{i'}\}_{i'\in\cT_u})\\
            = &H(\{Z_{i,j}:i\in\cU_j\backslash\cT_u\}_{j\in\cT_h}|\{Z_{i'}\}_{i'\in\cT_u})\\
            \geq &H(\{Z_{i,j}:i\in\cU_j\backslash\cT_u\}_{j\in\cT_h}|\{W_i:i\in\cU_j\backslash\cT_u\}_{j\in\cT_h},\{W_{i'},Z_{i'}\}_{i'\in\cT_u})\\
            \geq &I(\{Z_{i,j}:i\in\cU_j\backslash\cT_u\}_{j\in\cT_h};\{X_{i,j}:i\in\cU_j\backslash\cT_u\}_{j\in\cT_h}|\{W_i:i\in\cU_j\backslash\cT_u\}_{j\in\cT_h},\{W_{i'},Z_{i'}\}_{i'\in\cT_u})\\
            =& H(\{X_{i,j}:i\in\cU_j\backslash\cT_u\}_{j\in\cT_h}|\{W_i:i\in\cU_j\backslash\cT_u\}_{j\in\cT_h},\{W_{i'},Z_{i'}\}_{i'\in\cT_u})\nonumber\\&-H(\{X_{i,j}:i\in\cU_j\backslash\cT_u\}_{j\in\cT_h}|\{W_i:i\in\cU_j\backslash\cT_u\}_{j\in\cT_h},\{W_{i'},Z_{i'}\}_{i'\in\cT_u},\{Z_{i,j}:i\in\cU_j\backslash\cT_u\}_{j\in\cT_h})\\
            =&H(\{X_{i,j}:i\in\cU_j\backslash\cT_u\}_{j\in\cT_h}|\{W_i:i\in\cU_j\backslash\cT_u\}_{j\in\cT_h},\{W_{i'},Z_{i'}\}_{i'\in\cT_u})\\
            =&H(\{X_{i,j}:i\in\cU_j\backslash\cT_u\}_{j\in\cT_h}|\{W_{i'},Z_{i'}\}_{i'\in\cT_u})\nonumber\\&-I(\{X_{i,j}:i\in\cU_j\backslash\cT_u\}_{j\in\cT_h};\{W_i:i\in\cU_j\backslash\cT_u\}_{j\in\cT_h}|\{W_{i'},Z_{i'}\}_{i'\in\cT_u})\\
            =&H(\{X_{i,j}:i\in\cU_j\backslash\cT_u\}_{j\in\cT_h}|\{W_{i'},Z_{i'}\}_{i'\in\cT_u})\\
            \geq&\sum_{(i,j):j\in\cT_h,\ i\in\cU_j\backslash\cT_u}H(X_{i,j}|\{X_{i',j}\}_{i'\in\cU_j\backslash\cT_u,i'\neq i},\{W_{i'},Z_{i'}\}_{i'\in\cT_u})\\
            \geq&\sum_{(i,j):j\in\cT_h,\ i\in\cU_j\backslash\cT_u}H(X_{i,j}|\{W_{i'},Z_{i'}\}_{i'\in[N]\backslash\{i\}}),
        \end{align}
    \end{subequations}
    where (\ref{eq:Zij}b) follows from the fact that $Z_{i,j}$ is a function of $Z_i$, inequalities (\ref{eq:Zij}c)- (\ref{eq:Zij}e) are derived from the properties of conditional entropy and conditional mutual information, equation (\ref{eq:Zij}f) is derived from the fact that $X_{i,j}$ is generated by $W_{i}$ and $Z_{i,j}$. In (\ref{eq:Zij}g), the second term is zero, which is a consequence of the $(0,T_h,T_u)$-security property. Inequalities (\ref{eq:Zij}i) and (\ref{eq:Zij}j) are obtained by applying the chain rule and the properties of conditional entropy.
\end{pf}

We are now ready to derive the lower bound on $R_{Z_\Sigma}$ under the condition that $R_X$ and $R_Z$ are minimized.

\emph{Proof of $    R_{Z_{\Sigma}} \ge \min\left\{\frac{T_h(T_u+m)}{n},\frac{T_un+T_hm}{n}\right\}$}: 
Let $T_h$ and $T_u$ be positive integers satisfying $T_hm + T_u < N$. For every subset $\mathcal{T}_h \subseteq [K]$ with cardinality $|\mathcal{T}_h| = T_h$, we can find a subset $\mathcal{T}_u(\mathcal{T}_h) \subseteq [N]$ with $|\mathcal{T}_u(\mathcal{T}_h)| = T_u$ such that $\mathcal{T}_u(\mathcal{T}_h) \cap \mathcal{U}_{\mathcal{T}_h} = \emptyset$. By definition, we have
\begin{subequations}\label{eq:lowerboundRzsum}
    \begin{align}
        \binom{K}{T_h}\cdot L_{Z_\Sigma}\geq&\binom{K}{T_h}\cdot H(Z_{\Sigma})\\
        \geq&\sum_{\cT_h\subseteq[K]:|\cT_h|=T_h}H(Z_{\cT_u(\cT_h)},\{Z_{i,j}:i\in\cU_j\}_{j\in\cT_h})\\
        =&\sum_{\cT_h\subseteq[K]:|\cT_h|=T_h}\left(H(\{Z_{i,j}:i\in\cU_j\}_{j\in\cT_h}|Z_{\cT_u(\cT_h)})+H(Z_{\cT_u(\cT_h)})\right)\\
        \geq&\sum_{\cT_h\subseteq[K]:|\cT_h|=T_h}\left(\sum_{(i,j):j\in\cT_h,\ i\in\cU_j}H(X_{i,j}|\{W_{i'},Z_{i'}\}_{i'\neq i})+T_u\cdot\min\left\{L,\frac{T_hL}{n}\right\}\right)\\
        \geq&\binom{K}{T_h}T_u\cdot\min\left\{L,\frac{T_hL}{n}\right\}+\binom{K-1}{T_h-1}\sum_{i\in[N]}\sum_{j\in\cH_i}H(X_{i,j}|\{W_{i'},Z_{i'}\}_{i'\neq i})\\
        \geq&\binom{K}{T_h}T_u\cdot\min\left\{L,\frac{T_hL}{n}\right\}+\binom{K-1}{T_h-1}\sum_{i\in[N]}H(\{X_{i,j}\}_{j\in\cH_i}|\{W_{i'},Z_{i'}\}_{i'\neq i})\\
        \geq&\binom{K}{T_h}T_u\cdot\min\left\{L,\frac{T_hL}{n}\right\}+\binom{K-1}{T_h-1}NL,
    \end{align}
\end{subequations}
where (\ref{eq:lowerboundRzsum}b) holds because $Z_{i}$ and $Z_{i,j}$ are generated by $Z_{\Sigma}$. Inequality (\ref{eq:lowerboundRzsum}d) follows from Lemma~\ref{lem:sizeofZTu} and Lemma~\ref{lem:ZijtoXij}.  In (\ref{eq:lowerboundRzsum}e), each pair $(i,j)$ appears exactly $\binom{K-1}{T_h-1}$ times within the summation. Finally, (\ref{eq:lowerboundRzsum}g) is a direct consequence of Lemma~\ref{lem:xijyj}.

Dividing both sides of (\ref{eq:lowerboundRzsum}) by $\binom{K}{T_h}$, we obtain
\begin{align*}
    L_{Z_\Sigma} &\geq T_u\cdot\min\left\{L,\frac{T_hL}{n}\right\} + \frac{T_h NL}{K} \\
    &= T_u\cdot\min\left\{L,\frac{T_hL}{n}\right\} + \frac{T_h m L}{n} \\&= \min\left\{\frac{T_un+T_hm}{n}L,\frac{T_uT_h+T_hm}{n}L\right\},
\end{align*}
which directly implies that $R_{Z_{\Sigma}} \geq \min\{(T_un+T_hm)/n,(T_uT_h+T_hm)/n\}$.

\section{Achievable schemes}\label{sec:scheme}
In this section, we present three distinct constructions of secure aggregation schemes in Theorem~\ref{thm:scheme(1,u-1)}, Theorem~\ref{thm:scheme(1/n)}, and Theorem~\ref{thm:caseT=N-3}, respectively. 
\subsection{Scheme with optimal communication load and large key rates}
The first scheme achieves a key rate of $(R_Z, R_{Z_\Sigma}) = (1, N-1)$ and attains the optimal communication load for any homogeneous user-relay connection. The construction leverages concepts from network function computation theory. Specifically, the scheme proceeds in two primary steps:

\begin{itemize}
    \item \textbf{Step 1}. We begin by establishing a non-secure network code $\mathcal{C}$ with rate $n$ that is capable of computing the algebraic sum over the corresponding three-layer network based on Theorem~\ref{thm:NFC_cutset}.
    \item \textbf{Step 2}. Building upon $\mathcal{C}$, we introduce a random key associated with each link between the first two layers. A total of $Nn$ random keys are employed, subject to $n$ linear constraints, resulting in $(N-1)n$ degrees of freedom. Finally, we design the key generation process to ensure security.
\end{itemize}
The core idea of this scheme is to maximize its robustness against collusion by introducing a sufficient amount of random keys into the system. In this design, each link carries a key of entropy $H(Z_{i,j})=L/n$. Since each user is connected to $n$ relays, the total key entropy per user becomes $H(Z_i)=L$, which is sufficient for the security requirement. Meanwhile, to maintain the decodability, the random keys $Z_1, \cdots, Z_N$ must be linearly dependent, leading to a total key entropy of $R_{Z_{\Sigma}}=N-1$. This scheme is, therefore, advantageous when communication load and collusion resistance are prioritized over the storage cost of random keys.
In the following,  we provide a detailed description of the proposed scheme.

In Step 1, we define $\cN=(\cV,\cE)$ as a three-layer network, where users are treated as source nodes and relays as middle nodes. Consequently, the network comprises $N$ source nodes, $K$ middle nodes and a single sink node. The minimum cut size in $\cN$ is $n$. Based on Theorem~\ref{thm:NFC_cutset}, we construct a linear network code of rate $n$ that computes the sum function over $\cN$. 
In this code, each $W_i$ is divided into $n$ subsequences of length $L/n$. 
Without loss of generality, we may treat each block of symbols from $W_k$ as an element over an extension field. Specifically, $W_i = (W_i^{(1)}, \dots, W_i^{(n)})$ is a random vector of length $n$ over $\mathbb{F}_{q^{L/n}}$. For simplicity, we still denote the field as $\mathbb{F}_q$, and every element of matrices constructed in the following is in the larger field.
The network code is specified by the following four matrices, as defined in Section~\ref{sec:preliminary}: $\textbf{D} \in \mathbb{F}_q^{n \times K}$, $\textbf{E}_i \in \mathbb{F}_q^{n \times n}$, $\textbf{P} \in \mathbb{F}_q^{|\mathcal{E}| \times |\mathcal{E}|}$, and $\textbf{A}_{\rho} \in \mathbb{F}_q^{|\rho| \times |\mathcal{E}|}$. Their specific choices are as follows.

Let $\mathbf{d}_j \in \mathbb{F}_q^{n \times 1}$ denote the $j$-th column of $\textbf{D}$, and let $\mathcal{H}_i \subseteq [K]$ represent the set of middle nodes connected to source node $i$. The matrix $\textbf{D}$ must satisfy the condition that any submatrix formed by the columns indexed by $\mathcal{H}_i$ is full rank. That is, for every $i \in [N]$,
\begin{equation}\label{eq:defDi}
    \textbf{D}_i\eqdef(\bd_j:j\in\cH_i)\in\mathbb{F}_q^{n\times n}
\end{equation}
is full rank. We then define $E_i$ as its inverse,
\begin{equation}\label{eq:defEi}
    \textbf{E}_i\eqdef \textbf{D}_i^{-1}.
\end{equation} 
For a subset of links $\rho \subseteq \cE$, let $\textbf{A}_\rho=(a_{d,e})_{d \in \rho, e\in \cE}$ where
\begin{equation*}
  a_{d,e}=
  \begin{cases}
    1, & \mbox{if $d=e$}; \\
    0, & \mbox{otherwise}.
  \end{cases}
\end{equation*}
Let $\textbf{P}=(p_{d,e})_{d \in \mathcal{E}, e\in \mathcal{E}}$ such that
\begin{equation*}
  p_{d,e}=
  \begin{cases}
    1, & \mbox{if $\head(d)=\tail(e)$}; \\
    0, & \mbox{otherwise}.
  \end{cases}
\end{equation*}
The non-secure network code $\mathcal{C}$ can then be described as follows. Each source node $i$ possesses a message $\mathbf{x}_i \in \mathbb{F}_q^{1 \times n}$ and transmits $\mathbf{x}_i \cdot \textbf{E}_i^{T}$. Specifically, for any $j \in \mathcal{H}_i \subseteq [K]$, source node $i$ sends $u_{i,j} = (\mathbf{x}_i \cdot \textbf{E}_i^{T})_j \in \mathbb{F}_q$ through the link $ij$. It is important to note that $\mathbf{x}_i \cdot \textbf{E}_i^{T}$ is a vector of length $n$, whose coordinates are indexed by $\mathcal{H}_i$ rather than $[n]$; thus, $(\mathbf{x}_i \cdot \textbf{E}_i^{T})_j$ refers to the coordinate corresponding to index $j$. Subsequently, each middle node simply sums all received messages and transmits the result to the sink node. That is, middle node $j$ sends $u_j = \sum_{i \in \mathcal{U}_j} u_{i,j}$ to the sink node.
Thus, the sink node receives
\begin{equation}
    \bx\cdot\begin{pmatrix}
        \textbf{E}_1^T\cdot \textbf{A}_{Out(1)}\\
        \textbf{E}_2^T\cdot \textbf{A}_{Out(2)}\\
        \vdots\\
        \textbf{E}_U^T\cdot \textbf{A}_{Out(U)}
    \end{pmatrix}\cdot (\textbf{I}-\textbf{P})^{-1}\cdot \textbf{A}_{In(\gamma)}^T\in\mathbb{F}_q^{1\times K}.
\end{equation}
Based on the results in network function computation, the sink node can finish the computation by multiplying the decoding matrix $D^T$ at the right-hand side, i.e.,
\begin{align*}
    &\bx\cdot\begin{pmatrix}
        \textbf{E}_1^T & 0&\cdots&0\\
        0&\textbf{E}_2^T&\cdots&0\\
        \vdots&\vdots&\ddots&\vdots\\
        0&0&\cdots&\textbf{E}_U^T
    \end{pmatrix}
    \begin{pmatrix}
        \textbf{A}_{Out(1)}\\
        \textbf{A}_{Out(2)}\\
        \vdots\\
        \textbf{A}_{Out(U)}
    \end{pmatrix}\cdot(\textbf{I}-\textbf{P})^{-1}\cdot \textbf{A}_{In(\gamma)}^T\cdot \textbf{D}^T\\
    =&\bx\cdot\begin{pmatrix}
        \textbf{E}_1^T & 0&\cdots&0\\
        0&\textbf{E}_2^T&\cdots&0\\
        \vdots&\vdots&\ddots&\vdots\\
        0&0&\cdots&\textbf{E}_U^T
    \end{pmatrix}\begin{pmatrix}
        \textbf{D}_1^T\\
        \textbf{D}_2^T\\
        \vdots\\
        \textbf{D}_U^T
    \end{pmatrix}=\sum_{i=1}^N \bx_i.
\end{align*}
In particular, we can select $\textbf{D}$ as a maximum distance separable (MDS) matrix, then the $\textbf{D}_i,\textbf{E}_i$ are determined by the $\textbf{D}$ due to (\ref{eq:defDi}) and $(\ref{eq:defEi})$. 

In Step 2, each source node $i$ is assigned with a random vector $Z_i \in \mathbb{F}_q^{1 \times n}$, whose coordinates are indexed by $\mathcal{H}_i$, i.e., $Z_i = (Z_{i,j}: j \in \mathcal{H}_i)$. Let $\{R_i: i \in [(N-1)n]\}$ denote a set of $(N-1)n$ independent and identically distributed (i.i.d.) random variables drawn uniformly from $\mathbb{F}_q$.\footnote{In this scheme, the key generation is independent of the values of $T_u$ and $T_h$. This is because that to guarantee security under the maximum collusion size, we use a sufficient amount of randomness. In other words, this scheme is $(0,K-n,n(\cN_{N.K,n},K-n)-1)$-secure, therefore, it is also $(0,T_h,T_u)$-secure for any $T_h\leq K-n$ and $T_u\leq n(\cN_{N.K,n},K-n)-1$.} For $i \in [N-1]$, we define $Z_{i,j} = R_{(i-1)n+j}$, which implies $Z_1 = (R_1, \dots, R_n), \dots, Z_{N-1} = (R_{(N-2)n+1}, \dots, R_{(N-1)n})$. Furthermore, the vectors $Z_1, \dots, Z_N$ are constrained by
\begin{equation}\label{eq:Z_constraint}
    \begin{pmatrix}
        Z_1&Z_2&\cdots&Z_N
    \end{pmatrix}\cdot
    \begin{pmatrix}
        \textbf{D}_1^T\\
        \textbf{D}_2^T\\
        \vdots\\
        \textbf{D}_N^T
    \end{pmatrix}=0,
\end{equation}
where $\textbf{D}_i$ is defined in (\ref{eq:defDi}). Consequently, $Z_N$ can be expressed as
\begin{equation}
    Z_N=\sum_{i\in[N-1]}Z_i\cdot \textbf{D}_i^T\cdot(\textbf{D}_N^T)^{-1}=(R_1,\cdots,R_{(N-1)n})\cdot\begin{pmatrix}
        \textbf{D}_1^T(\textbf{D}_N^T)^{-1}\\
        \textbf{D}_2^T(\textbf{D}_N^T)^{-1}\\
        \vdots\\
        \textbf{D}_{N-1}^T(\textbf{D}_N^T)^{-1}
    \end{pmatrix}.
\end{equation}
Within the HSA scheme, for $i \in [N]$ and $j \in \mathcal{H}_i$, each user $i$ transmits the message $X_{i,j} = (W_i \textbf{D}_i^T)_j + Z_{i,j}$. Each relay aggregates all the received messages and forwards the sum to the server, i.e., $Y_j = \sum_{i \in \mathcal{U}_j} X_{i,j}$.

Before verifying the correctness of the proposed scheme, we present an illustrative example for clarity.

\begin{figure}
    \centering
    \includegraphics[width=0.65\linewidth]{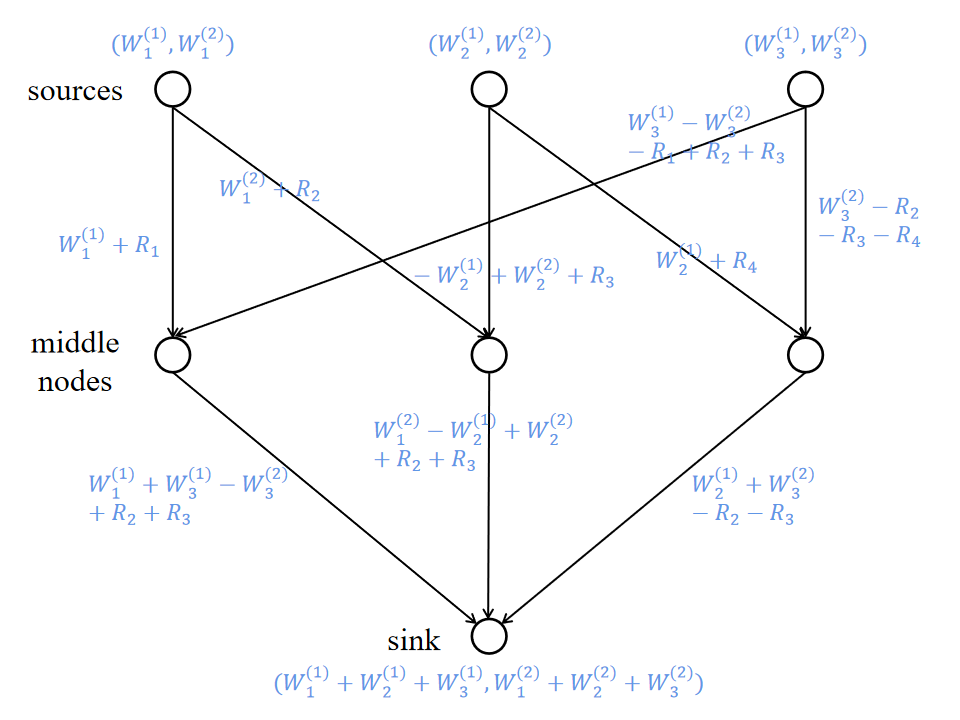}
    \caption{A hierarchical secure aggregation model with $3$ users and $3$ relays.}
    \label{fig:example_3,3,2,2_}
\end{figure}

\begin{example}\label{eg:example_3,3,2,2_}
    Consider an HSA problem characterized by the parameters $(N, K, n, T_h, T_u) = (3, 3, 2, 1, 1)$, as illustrated in Fig.~\ref{fig:example_3,3,2,2_}. We select the decoding matrix $\textbf{D}$ as
    \[\textbf{D}=\begin{pmatrix}
        1&0&1\\
        0&1&1
    \end{pmatrix}.\]
    This ensures that $E$ satisfies the MDS property. Consequently, the matrices $D_1, D_2,$ and $D_3$ are given by 
    \[\textbf{D}_1=\begin{pmatrix}
        1&0\\
        0&1
    \end{pmatrix},\ \textbf{D}_2=\begin{pmatrix}
        0&1\\
        1&1
    \end{pmatrix},\ \textbf{D}_3=\begin{pmatrix}
        1&1\\
        0&1
    \end{pmatrix}.\]
    This leads to the following source encoding matrices $\textbf{E}_1, \textbf{E}_2,$ and $\textbf{E}_3$:
    \[\textbf{E}_1=\begin{pmatrix}
        1&0\\
        0&1
    \end{pmatrix},\ \textbf{E}_2=\begin{pmatrix}
        -1&1\\
        1&0
    \end{pmatrix},\ \textbf{E}_3=\begin{pmatrix}
        1&-1\\
        0&1
    \end{pmatrix}.\]
    The random key assignment is then defined as
    \begin{align*}
       Z_1=(R_1,R_2),\ \ Z_2=(R_3,R_4),\ \  Z_3=(-R_1+R_2+R_3,-R_2-R_3-R_4).
    \end{align*}
    It is straightforward to verify that the constraint in (\ref{eq:Z_constraint}) is satisfied.  The message transmitted from user $1$ to relay $j$ can be computed as
    \[X_{1,1}=((W_1^{(1)},W_{1}^{(2)})\cdot E_1^T)_1+Z_{1,1}=W_1^{(1)}+R_1.\]
    Similarly, the remaining messages $X_{i,j}$ are:
    \begin{align*}
    &X_{1,2}=W_{1}^{(2)}+R_2,\\
    &X_{2,2}=-W_2^{(1)}+W_2^{(2)}+R_3,\\
    &X_{2,3}=W_{2}^{(1)}+R_4,\\
    &X_{3,1}=-W_{3}^{(1)}+W_{3}^{(2)}-R_1+R_2+R_3,\\
    &X_{3,3}=W_{3}^{(1)}-R_2-R_3-R_4.\\
\end{align*}
Each relay then sums the messages it receives and transmits the result:
\begin{align*}
    &Y_{1}=W_{1}^{(1)}+W_3^{(1)}-W_3^{(2)}+R_2+R_3,\\
    &Y_{2}=W_{1}^{(2)}-W_{2}^{(1)}+W_2^{(2)}+R_2+R_3,\\
    &Y_{3}=W_{2}^{(1)}+W_{3}^{(2)}-R_2-R_3.
\end{align*}
The server can then decode $W_1 + W_2 + W_3$ from the sums $Y_1 + Y_3$ and $Y_2 + Y_3$.  This scheme achieves $(0, 1, 1)$-security. For instance, consider the scenario where $\mathcal{T}_u = \{1\}$ and $\mathcal{T}_h = \{3\}$, we have
\begin{subequations}\label{seq:example3322}
    \begin{align}
        &I(X_{2,3},X_{3,3};W_{[3]}|W_1,Z_1)\\
        =&H(X_{2,3},X_{3,3}|W_1,Z_1)-H(X_{2,3},X_{3,3}|W_{[3]},Z_1)\\
        =&H(X_{2,3},X_{3,3}|W_1,Z_1)-H(R_3,-R_3-R_4)\\
        \leq &H(X_{2,3})+H(X_{3,3})-2=0,
    \end{align}
\end{subequations}
where (\ref{seq:example3322}b) is obtained by substituting $X_{2,3}$, $X_{3,3}$, and $Z_1$ into the second term. This holds because $W_{[3]}$ is independent of $Z_{[3]}$, and $R_3, R_4$ are independent of $R_1, R_2$. The last inequality follows from the equalities $H(X_{i,j}) = H(R_i) = L/n = 1$.
The security arguments for other combinations of $\mathcal{T}_u$ and $\mathcal{T}_h$ are analogous.
\end{example}

In the following, we will demonstrate that for any integers $T_h$ and $T_u$ satisfying $T_h \leq K-n$ and $T_u \leq n(\mathcal{N}_{N,K,n}, T_h) - 1$, the proposed HSA scheme is both decodable and $(0, T_h, T_u)$-secure.

\emph{\textbf{Decodability}}: Within the scheme, each user $i$ transmits the message
\[X_{i,j}=\left((W_i,Z_i)\cdot\begin{pmatrix}
    \textbf{E}_i^T\\
    \textbf{I}_{n}
\end{pmatrix}\right)_j\]
to the relay $j$, where $\textbf{I}_n$ is an $n\times n$ identity matrix, and $(\cdot)_j$ is the $j$-th coordinate of the vector in parentheses. Consequently, the server receives
\begin{equation}
    (W_1,Z_1,\cdots,W_N,Z_N)\cdot\begin{pmatrix}
        \textbf{E}_1' & 0&\cdots&0\\
        0&\textbf{E}_2'&\cdots&0\\
        \vdots&\vdots&\ddots&\vdots\\
        0&0&\cdots&\textbf{E}_N'
    \end{pmatrix}
    \begin{pmatrix}
        \textbf{A}_{Out(1)}\\
        \textbf{A}_{Out(2)}\\
        \vdots\\
        \textbf{A}_{Out(U)}
    \end{pmatrix}\cdot(\textbf{I}-\textbf{P})^{-1}\cdot \textbf{A}_{In(\gamma)}^T
\end{equation}
where $\textbf{E}_i'=\begin{pmatrix}\textbf{E}_i&\textbf{I}_{n}\end{pmatrix}^T$. The server then multiplies the received data by $\textbf{D}^T$ on the right-hand side, leading to:
\begin{subequations}
    \begin{align}
        &(W_1,Z_1,\cdots,W_N,Z_N)\cdot\begin{pmatrix}
        \textbf{E}_1' & 0&\cdots&0\\
        0&\textbf{E}_2'&\cdots&0\\
        \vdots&\vdots&\ddots&\vdots\\
        0&0&\cdots&\textbf{E}_N'
        \end{pmatrix}\begin{pmatrix}
        \textbf{D}_1^T\\
        \textbf{D}_2^T\\
        \vdots\\
        \textbf{D}_U^T
        \end{pmatrix}\\
        =&(W_1,Z_1,\cdots,W_N,Z_N)\cdot\begin{pmatrix}
            \textbf{I}_n\\
            \textbf{D}_1^T\\
            \vdots\\
            \textbf{I}_n\\
            \textbf{D}_N^T
        \end{pmatrix}\\
        =&\sum_{i=1}^N W_i+\sum_{i=1}^N Z_i \textbf{D}_i^T=\sum_{i=1}^N W_i.
    \end{align}
\end{subequations}
Therefore, the server is able to decode $\sum_{i=1}^N W_i$ from the set $\{Y_j : j \in [K]\}$.

\emph{$\bf(0,T_h,T_u)$\textbf{-security}}: 
In the proposed HSA scheme, the user $i$ transmits $X_{i,j}=(W_i\textbf{D}_i^T)_j+Z_{i,j}$ to the relay $j$. Consequently, for every relay $j\in[K]$, denote $\cU_j=\{i_1,\cdots,i_m\}$, then the messages received by $j$ can be represented as
\begin{equation}
    \begin{pmatrix}
        X_{i_1,j}\\\vdots\\X_{i_m,j}
    \end{pmatrix}\eqdef \textbf{C}_jW^T+\textbf{M}_jZ^T,
\end{equation}
where $\textbf{C}_j\in\mathbb{F}_q^{m\times Nn},\textbf{M}_j\in\{0,1\}^{m\times Nn}$ are coefficient matrices about the source messages and random keys, $W=(W_1,\cdots,W_N)$, $Z=(Z_1,\cdots,Z_N)$ are the vectors of source messages and random keys. The columns of $\textbf{M}_j$ can be partitioned into $N$ blocks, with each block corresponding to a specific user $i$ and its respective $Z_i$. Furthermore, within the block corresponding to user $i$, each column is indexed by a symbol in $\cH_i$, and only the column indexed by $j$ contains non-zero entries. For instance, in Example~\ref{eg:example_3,3,2,2_}, 
\[\textbf{M}_1=\begin{pmatrix}
    1&0&0&0&0&0\\
    0&0&1&0&0&0
\end{pmatrix},\]
where the first two columns correspond to user $1$, and specifically, the first column in the first block corresponds to relay $1 \in \mathcal{H}_1$.
For any subset $\mathcal{S} \subseteq [N]$, we define $\textbf{M}_j(\cS) \in \mathbb{F}_q^{m \times |\mathcal{S}|n}$ as a submatrix of $\textbf{M}_j$, formed by selecting only the columns corresponding to the users in $\cS$. Finally, $\textbf{M}_{\mathcal{T}_h}(\cS)$ is defined as the vertical concatenation of $\textbf{M}_j(\cS)$ for all $j \in \mathcal{T}_h$. In Example~\ref{eg:example_3,3,2,2_}, we have 
\[\textbf{M}_1(\{1,2\})=\begin{pmatrix}
    1&0&0&0\\
    0&0&1&0
\end{pmatrix}.\]

Next, we define $\textbf{D}_Z \eqdef \begin{pmatrix} \textbf{D}_1 & \textbf{D}_2 & \cdots & \textbf{D}_N \end{pmatrix}$.  Similarly to the previous notations, the columns of $\textbf{D}_Z$ are partitioned into $N$ blocks, each block associated with a specific user $i$.  We denote $\textbf{D}_Z(\cS)$ as the submatrix of $\textbf{D}_Z$ comprised of the blocks corresponding to the users within the set $\cS$.  For instance, in Example~\ref{eg:example_3,3,2,2_}, we have
\[\textbf{D}_Z(\{1,2\})=\begin{pmatrix}
    \textbf{D}_1&\textbf{D}_2
\end{pmatrix}=\begin{pmatrix}
    1&1&1&0\\
    0&1&0&1
\end{pmatrix}.\]

Let $\mathcal{T}_h$ and $\mathcal{T}_u$, where $|\cT_h|=T_h,|\cT_u|=T_u$, denote the sets of colluding relays and users, respectively. Based on the constructed HSA scheme and the definitions of $\textbf{M}_j,\textbf{M}_j(\cS),\textbf{M}_{\mathcal{T}_h}(\cS)$, we have the following properties.
\begin{prop}\label{prop:propertyofM}
    For every $j\in[K]$, the following statements hold:
    \begin{enumerate}
        \item A column of $\textbf{M}_j(\bar{\mathcal{T}}_u)$ is a zero vector if and only if it corresponds to a block associated with a user $i\in\bar{\mathcal{T}}_u$ and is labeled by a symbol $j' \in \mathcal{H}_i$, where $j' \neq j$;
        \item Each row of $\textbf{M}_{\mathcal{T}_h}(\cS)$ contains exactly one `1', and each column of $\textbf{M}_{\mathcal{T}_h}(\cS)$ contains at most one `1'.
    \end{enumerate}
\end{prop}
\begin{pf}
    According to the definition of $\textbf{M}_j(\bar{\mathcal{T}}_u)$, its columns are partitioned into $N$ blocks, each corresponding to a user in $\bar{\mathcal{T}}_u$. Within each block, columns are indexed by the symbols in $\cH_i$, and only the column indexed by $j$ contains non-zero entries. Therefore, statement 1) follows directly from the definition.
    For 2), recall that in the constructed HSA scheme, $X_{i,j} = (W_i\textbf{D}_i^T)_j + Z_{i,j}$. Since each $Z_{i,j}$ appears exclusively in $X_{i,j}$, the matrix $[\textbf{M}_1; \textbf{M}_2; \cdots; \textbf{M}_N]$ becomes an identity matrix after an appropriate row permutation. As $\textbf{M}_{\mathcal{T}_h}(\cS)$ is a submatrix of $[\textbf{M}_1; \textbf{M}_2; \cdots; \textbf{M}_N]$, statement 2) holds.
\end{pf}
 

The $(0,T_h,T_u)$-security is established as follows. First, we prove that $\mathcal{M}_{\mathcal{T}_h}(\bar{\mathcal{T}}_u) \cap \mathcal{D}_Z(\bar{\mathcal{T}}_u) = {\mathbf{0}}$. Then, we derive the security from this condition. Formally, we have
\begin{lem}\label{lem:securitypf_1}
    If $0<T_h\leq K-n,T_u\leq n(\cN_{N,K,n},T_h)-1$, then
    \begin{equation}\label{eq:lem_spf_1}
        \mathcal{M}_{\mathcal{T}_h}(\bar{\mathcal{T}}_u) \cap \mathcal{D}_Z(\bar{\mathcal{T}}_u) = \{\boldsymbol{0}\},
    \end{equation}
    where $\mathcal{M}_{\mathcal{T}_h}(\bar{\mathcal{T}}_u)$ and $\mathcal{D}_Z(\bar{\mathcal{T}}_u)$ are the row spaces spanned by $\textbf{M}_{\mathcal{T}_h}(\bar{\mathcal{T}}_u)$ and $\textbf{D}_Z(\bar{\mathcal{T}}_u)$, respectively.
\end{lem}
\begin{figure}
    \centering
    \includegraphics[width=0.8\linewidth]{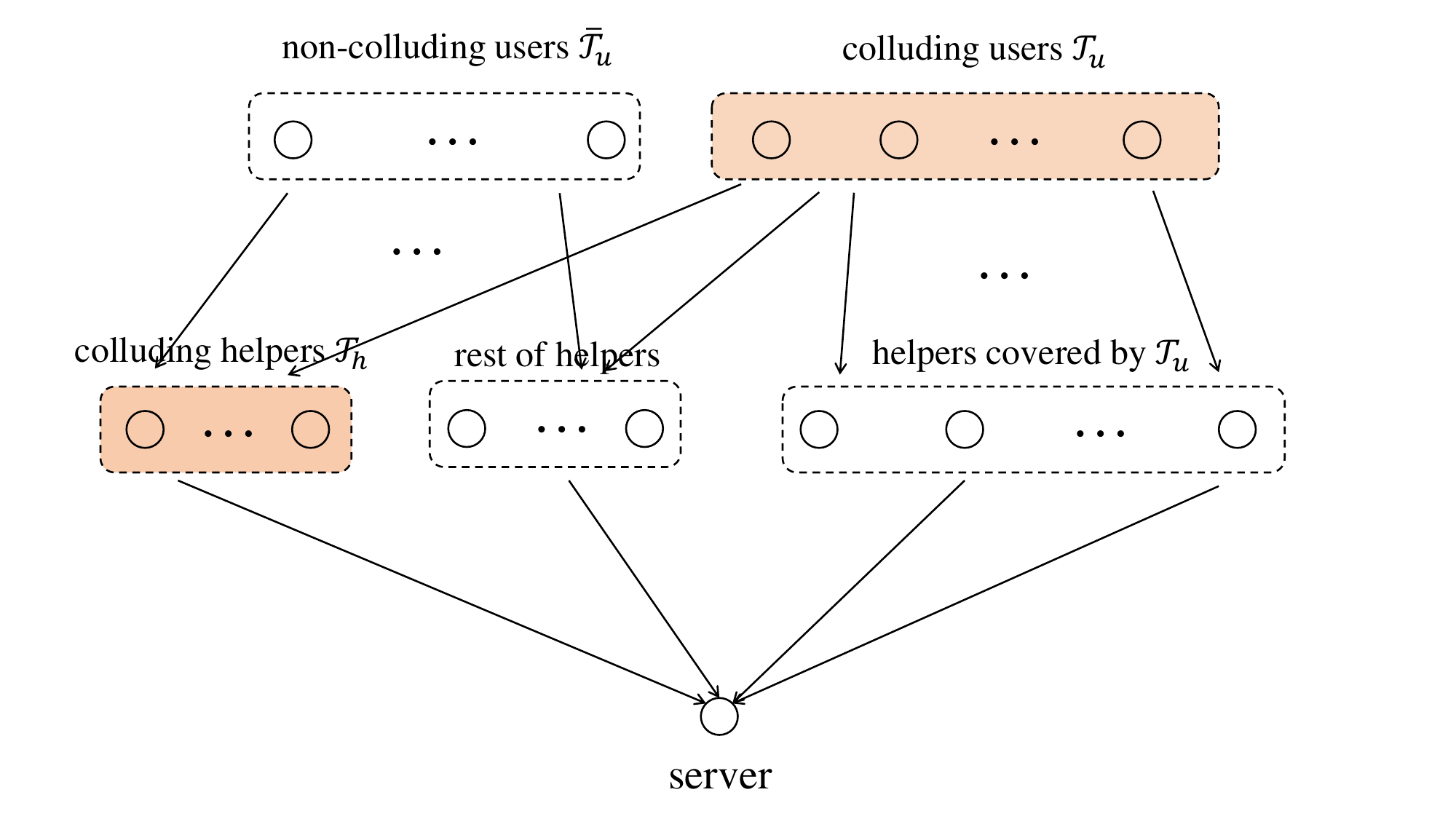}
    \caption{An illustration of Lemma~\ref{lem:securitypf_1}.}
    \label{fig:illustration}
\end{figure}
\begin{pf}
    From the first statement in Proposition~\ref{prop:propertyofM}, $M_{\mathcal{T}_h}(\bar{\mathcal{T}}_u)$ contains $n$ columns that correspond to $n$ distinct symbols ${j_1, j_2, \ldots, j_n}$ and form an all-zero submatrix if and only if there exist $n$ distinct relays ${j_1, j_2, \ldots, j_n}$ connected exclusively to the non-colluding users in $\bar{\mathcal{T}}_u$.

    Recall that $n(\mathcal{N}_{N,K,n}, T_h)$ denotes the minimum number of users required to cover exactly $K-T_h-n+1$ relays. Given that $T_u \leq n(\mathcal{N}_{N,K,n}, T_h) - 1$, the colluding users can cover at most $K-T_h-n$ relays. Therefore, the remaining relays must be connected solely to users in $\bar{\mathcal{T}}_u$. Excluding those in $\mathcal{T}_h$, at least $n$ relays are connected to at least one user in $\bar{\mathcal{T}}_u$, as illustrated in Fig.~\ref{fig:illustration}. Hence, the required set of $n$ columns indeed exists.

    We can thus identify $n$ columns in $\textbf{M}_{\cT_h}(\bar{\cT}_u)$ that form an all-zero submatrix. Meanwhile, the corresponding columns in $\textbf{D}_Z(\bar{\cT}_u)$, given by $\{\bd_j: j \in {j_1,j_2,\cdots,j_n}\}$, are full rank because $\textbf{D}$ is an MDS matrix. Consequently, no linear combination of the rows of $\textbf{D}_Z(\bar{\mathcal{T}}u)$ can produce a matrix whose columns corresponding to ${j_1,j_2,\cdots,j_n}$ are all-zero. This implies that $\mathcal{M}_{\mathcal{T}_h}(\bar{\mathcal{T}}_u) \cap \mathcal{D}_Z(\bar{\mathcal{T}}_u) = {\boldsymbol{0}}$.
\end{pf}

In Example~\ref{eg:example_3,3,2,2_}, where $\mathcal{T}_u = \{3\}$ and $\mathcal{T}_h = \{1\}$, we have
\[\textbf{M}_1(\{1,2\})=\begin{pmatrix}
    1&0&0&0\\
    0&0&1&0
\end{pmatrix},\ \textbf{D}_Z(\{1,2\})=\begin{pmatrix}
    1&1&1&0\\
    0&1&0&1
\end{pmatrix}.\]
It is evident that the submatrix of $\textbf{M}_1(\{1,2\})$ formed by the second and fourth columns is a zero matrix, while the corresponding submatrix of $\textbf{D}_Z(\{1,2\})$ is full rank. Consequently, the row spaces of $\textbf{M}_1(\{1,2\})$ and $\textbf{D}_Z(\{1,2\})$ indeed intersect trivially.

Now, it suffices to prove the security based on (\ref{eq:lem_spf_1}).
\begin{lem}\label{lem:securitypf_2}
    If $\mathcal{M}_{\cT_h}(\bar{\cT}_u)\cap\cD_Z(\bar{\cT}_u)=\{\Zero\}$, then we have $I(W_{[N]};\{X_{i,j}:i\in\cU_j\}_{j\in\cT_h}|\{W_i,Z_i\}_{i\in\cT_u})=0$.
\end{lem}
\begin{pf}
    In the constructed scheme, the random keys $Z_1,\cdots,Z_N$ satisfy (\ref{eq:Z_constraint}), which leads to the following results,
    \begin{align}
       &H(Z_{[N]})=(N-1)n,\\
       &H(Z_{\cT_u})=T_un,\\
       &H(Z_{\bar{\cT}_u})=(N-T_u)n.
    \end{align}
    Consequently, the conditional entropy is $H(Z_{\bar{\cT}u}|Z_{\cT_u})=(N-T_u-1)n$. This implies that, given $Z_{\cT_u}$ and letting $\bc=\textbf{D}_Z(\cT_u)\cdot Z_{\cT_u}^T$, the variable $Z_{\bar{\cT}_u}$ is uniformly distributed over the affine space $\cW$ defined by \[\cW\eqdef\{\bx\in\mathbb{F}_q^{(N-T_u)n}:\textbf{D}_Z(\bar{\cT}_u)\bx^T=\bc\}.\]
   It follows that, given $Z_{\cT_u}$, the product $\textbf{M}_{\cT_h}(\bar{\cT}_u)\cdot Z_{\bar{\cT}_u}$ is uniformly distributed over the affine space $\textbf{M}_{\cT_h}(\bar{\cT}_u)\cW$, where
    \[\textbf{M}_{\cT_h}(\bar{\cT}_u)\cW\eqdef\{\textbf{M}_{\cT_h}(\bar{\cT}_u)\bx^T\in\mathbb{F}_q^{T_hm}:\bx\in\cW\}.\]
    Therefore, $H(\textbf{M}_{\cT_h}(\bar{\cT}_u)Z_{\bar{\cT}_u}^T|Z_{\cT_u})=\log_q|\textbf{M}_{\cT_h}(\bar{\cT}_u)\cW|$. Now, define the linear space $\cV\eqdef\{\bx\in\mathbb{F}_q^{(N-T_u)n}:\textbf{D}_Z(\bar{\cT}_u)\bx^T=\Zero\}.$ Since $\mathcal{M}_{\cT_h}(\bar{\cT}_u)\cap\cD_Z(\bar{\cT}_u)=\{\Zero\}$ and $\Rank(\textbf{M}_{\cT_h}(\bar{\cT}_u))=T_hm$, we have
    \begin{align*}
        \log_q|\textbf{M}_{\cT_h}(\bar{\cT}_u)\cW|&=\dim(\textbf{M}_{\cT_h}(\bar{\cT}_u)\cV)=\dim(\cV)-\dim(\cV\cap \rm{Null}(\textbf{M}_{\cT_h}(\bar{\cT}_u)))\\
        &=((N-T_u)n-n)-((N-T_u)n-n-T_hm)\\
        &=T_hm.
    \end{align*}
    Thus, we have
    \begin{equation}\label{eq:spf_2_1}
        H(\textbf{M}_{\cT_h}(\bar{\cT}_u)Z_{\bar{\cT}_u}^T|Z_{\cT_u})=\log_q|\textbf{M}_{\cT_h}(\bar{\cT}_u)\cW|=T_hm.
    \end{equation}
    Consequently, we have
    \begin{subequations}\label{eq:spf_2}
        \begin{align}
            &I(W_{[N]};\{X_{i,j}:i\in\cU_j\}_{j\in\cT_h}|\{W_i,Z_i\}_{i\in\cT_u})\\
            =&H(\{X_{i,j}:i\in\cU_j\}_{j\in\cT_h}|\{W_i,Z_i\}_{i\in\cT_u})-H(\{X_{i,j}:i\in\cU_j\}_{j\in\cT_h}|W_{[N]},\{Z_i\}_{i\in\cT_u})\\
            \leq&\sum_{j\in\cT_h,i\in\cU_j}H(X_{i,j})-H(\{X_{i,j}:i\in\cU_j\}_{j\in\cT_h}|W_{[N]},\{Z_i\}_{i\in\cT_u})\\
            \leq&T_hm-H(\{X_{i,j}:i\in\cU_j\}_{j\in\cT_h}|W_{[N]},\{Z_i\}_{i\in\cT_u})\\
            =&T_hm-H(\textbf{C}_{\cT_h}W^T+\textbf{M}_{\cT_h}Z^T|W,Z_{\cT_u})\\
            =&T_hm-H(\textbf{M}_{\cT_h}Z^T|W,Z_{\cT_u})\\
            =&T_hm-H(\textbf{M}_{\cT_h}(\bar{\cT}_u)Z_{\bar{\cT}_u}^T+\textbf{M}_{\cT_h}(\cT_u)Z_{\cT_u}^T|Z_{\cT_u})\\
            =&T_hm-H(\textbf{M}_{\cT_h}(\bar{\cT}_u)Z_{\bar{\cT}_u}^T|Z_{\cT_u})\\
            \overset{(\ref{eq:spf_2_1})}{=}&T_hm-T_hm=0,
        \end{align}
    \end{subequations}
    where (\ref{eq:spf_2}b) follows from the definition of mutual information; (\ref{eq:spf_2}c) uses the chain rule for conditional entropy; (\ref{eq:spf_2}d) holds since $X_{i,j}$ is a symbol over $\mathbb{F}_q$, which implies $H(X_{i,j})\leq1$; (\ref{eq:spf_2}e) uses the representation $\{X_{i,j}:i\in\cU_j\}_{j\in\cT_h}=\textbf{C}_{\cT_h}W^T+\textbf{M}_{\cT_h}Z^T$; (\ref{eq:spf_2}g) is due to the independence between $W$ and $Z$.
\end{pf}
From Lemma~\ref{lem:securitypf_1} and Lemma~\ref{lem:securitypf_2}, the $(0,T_h,T_u)$-security of the constructed scheme is proved.

\subsection{Scheme with optimal communication load and optimal key size}\label{subsec:scheme(1/n)}
When $T_h=1,T_u\leq n(\cN_{N,K,n},T_h)-1$, and $T_u+m< \min\{N-1,K-n\}$, for a multiple cyclic network, we can construct an HSA scheme with the communication load pair $(R_X,R_Y)=(1/n,1/n)$ and optimal key rate pair $(R_Z,R_{Z_\Sigma})=(\frac{1}{n},\frac{m+T_u}{n})$. 
In fact, the main idea of this scheme is to minimize the randomness stored per user. Since $H(Z_{i,j})\geq L/n$, we have $H(Z_i)\geq L/n$, which implies the random keys used in $\{X_{i,j}:j\in\cH_i\}$ are linearly dependent. For $R_{Z_\Sigma}$, since the colluding nodes can obtain at most $m+T_u$ messages, thus, the total random key size is at least $(m+T_u)L/n$. In this scheme, the encoding coefficients about the users' inputs $W_{[N]}$ are the same as the network code based on Theorem~\ref{thm:NFC_cutset}, while the coefficients about the random keys are carefully designed with respect to the network topology.
Formally, the scheme can be described as follows.

Each user $i \in [N]$ has an input $W_i = (W_i^{(1)}, \dots, W_i^{(n)}) \in \mathbb{F}_q^n$.
The system generates $T_u+m$ independent and uniformly distributed random keys over $\mathbb{F}_q$, denoted as $R_1, \dots, R_{T_u+m}$. Each user $i$ is assigned a single random key symbol, $Z_i$, generated from $(R_1, \dots, R_{T_u+m})$ via the following:
\begin{equation}\label{eq:s2_keygeneration}
    (Z_1,\cdots,Z_N)=(R_1\cdots,R_{T_u+m})\cdot \textbf{B},
\end{equation}
where $\textbf{B} \in \mathbb{F}_{q}^{(T_u+m) \times N}$ represents the key generating matrix, the specific form of which will be defined subsequently. Let $\textbf{D}$, $\textbf{D}_i$, and $\textbf{E}_i$ be the matrices defined in Equations (\ref{eq:defDi}) and (\ref{eq:defEi}).

In the first hop, each user $i \in [N]$ transmits 
\begin{equation}
    X_{i,j}=(W_i\textbf{E}_i^T)_j+\lambda_{i,j}Z_i
\end{equation}
to the relay $j$, $j\in\cH_i$.
In the second hop, each relay $j \in [K]$ sums the received messages and forwards
\[Y_j=\sum_{i\in\cU_j}X_{i,j}=\sum_{i\in\cU_j}(W_i\textbf{E}_i^T)_j+\sum_{i\in\cU_j}\lambda_{i,j}Z_i\]
to the server. Consequently, the server receives
\[(W_1,\cdots,W_N)\cdot\begin{pmatrix}
        \textbf{E}_1^T & 0&\cdots&0\\
        0&\textbf{E}_2^T&\cdots&0\\
        \vdots&\vdots&\ddots&\vdots\\
        0&0&\cdots&\textbf{E}_N^T
    \end{pmatrix}
    \begin{pmatrix}
        \textbf{A}_{Out(1)}\\
        \textbf{A}_{Out(2)}\\
        \vdots\\
        \textbf{A}_{Out(U)}
    \end{pmatrix}\cdot(\textbf{I}-\textbf{P})^{-1}\cdot \textbf{A}_{In(\gamma)}^T+(R_1,\cdots,R_{T_u+m})\cdot \textbf{B}\cdot{\bf{\Lambda}}.\]
where ${\bf{\Lambda}}=(\lambda_{i,j})_{i\in[N],j\in[K]}\in\mathbb{F}_q^{N\times K}$. If $j\notin\cH_i$, let $\lambda_{i,j}=0$. By right-multiplying by $\textbf{D}^T$, the server obtains
\begin{equation}\label{eq:s2_server}
    \sum_{i=1}^{N}W_i+(R_1,\cdots,R_{T_u+m})\cdot \textbf{B}\cdot{\bf{\Lambda}}\cdot \textbf{D}^T.
\end{equation}
Therefore, to ensure decodability, we must enforce
\begin{equation}\label{eq:BlambdaD=0}
    \textbf{B}\cdot{\bf{\Lambda}}\cdot \textbf{D}^T=\Zero_{(T_u+m)\times n}.
\end{equation}
Furthermore, for any $j\in [K]$ and $\mathcal{T}_u \subseteq [N]$, the colluding nodes can obtain at most $m + T_u$ random keys $Z_i$. Thus, if $\textbf{B}$ is an MDS matrix, then the $(0, 1, T_u)$-security is guaranteed.
In summary, we have the following lemma.

\begin{lem}
    If the matrices $\textbf{B}$, $\bf\Lambda$, and $\textbf{D}$ satisfy the following conditions:
    \begin{enumerate}[left=1pt]
        \item[(P1)] Both $\textbf{B}$ and $\textbf{D}$ are MDS matrices;
        \item[(P2)] For each user $i \in [N]$ and relay $j \notin \mathcal{H}_i$, $\lambda_{i,j} = 0$;
        \item[(P3)] $\textbf{B} \cdot {\bf{\Lambda}} \cdot \textbf{D}^T = \mathbf{0}_{(T_u+m) \times n}$,
    \end{enumerate}
    then, the proposed scheme is decodable and $(0,1,T_u)$-secure.
\end{lem}
\begin{pf}
    Since $\textbf{D}$ is an MDS matrix, then each $\textbf{D}_i,i\in[N]$ is full rank. From (\ref{eq:defDi}) and (\ref{eq:defEi}), if the server uses $\textbf{D}$ as the decoding matrix, then it obtains (\ref{eq:s2_server}). Then the decodabilitly is guaranteed due to (P3). 
    Since $B$ is an MDS matrix and based on the key generation equation (\ref{eq:s2_keygeneration}), any $T_u+m$ $Z_i$'s are independent.
    Therefore, for any two user subsets $\cS$ and $\cT_u$ with $|\cS|\leq m$ and $|\cT_u|=T_u$, we have
    \begin{equation}\label{eq:s2_lemmapf_1}
        H(Z_{\cS}|Z_{\cT_u})\leq|\cS\backslash\cT_u|.
    \end{equation} 
    For the $(0,1,T_u)$-security, for any relay $j\in[K]$ and colluding user set $\cT_u\subseteq[N],|\cT_u|=T_u$, we have
    \begin{subequations}\label{seq:s2_lemmapf}
        \begin{align}
            &I(\{X_{i,j}\}_{i\in\cU_j};W_{[N]}|\{W_i,Z_i\}_{i\in\cT_u})\\
            =&H(\{X_{i,j}\}_{i\in\cU_j}|\{W_i,Z_i\}_{i\in\cT_u})-H(\{X_{i,j}\}_{i\in\cU_j}|W_{[N]},\{Z_i\}_{i\in\cT_u})\\
            =&H(\{X_{i,j}\}_{i\in\cU_j}|\{W_i,Z_i\}_{i\in\cT_u})-H(\{\lambda_{i,j}Z_i\}_{i\in\cU_j}|\{Z_i\}_{i\in\cT_u})\\
            \overset{(\ref{eq:s2_lemmapf_1})}{=}&H(\{X_{i,j}\}_{i\in\cU_j}|\{W_i,Z_i\}_{i\in\cT_u})-|\cU_j\backslash\cT_u|\\
            \leq &\sum_{i\in\cU_j}H(X_{i,j}|\{W_i,Z_i\}_{i\in\cT_u})-|\cU_j\backslash\cT_u|\\
            \overset{(\ref{eq:first_hop_determinism})}{=}&\sum_{i\in\cU_j\backslash\cT_u}H(X_{i,j}|\{W_i,Z_i\}_{i\in\cT_u})-|\cU_j\backslash\cT_u|\\
            \leq &\sum_{i\in\cU_j\backslash\cT_u}H(X_{i,j})-|\cU_j\backslash\cT_u|
            \leq0,
        \end{align}
    \end{subequations}
    where (\ref{seq:s2_lemmapf}b) is due to the definition of mutual information, (\ref{seq:s2_lemmapf}c) is due to the independence between $W_{[N]}$ and $Z_{[N]}$, (\ref{seq:s2_lemmapf}e) and (\ref{seq:s2_lemmapf}g) are due to the property of conditional entropy, (\ref{seq:s2_lemmapf}g) is because that each $X_{i,j}$ is a single symbol over $\mathbb{F}_q$.
\end{pf}
It suffices to select the matrices $\textbf{B}$, $\bf{\Lambda}$, and $\textbf{D}$ satisfying (P1),(P2) and (P3).
For a multiple cyclic network, we consider the case where $N = tK$. 
The proposed scheme leverages the approach in \cite{2022ZhaoSun} by treating the blocks of sufficiently long input vectors $W_i$ (for $i\in[N]$) as elements in an extension field. This allows operations such as element wise
sum to be defined over the field extension, justified by the sufficient length of the vectors. Therefore, without loss of generality, we assume that the base field size $q$ is large enough.
We construct the matrices $\textbf{B}$, $\bf{\Lambda}$, and $\textbf{D}$ over $\mathbb{F}_q$ as follows.  
We partition the network into $t$ cyclic subnetworks. Consequently, we can partition $\textbf{B}$ and $\bf{\Lambda}$ into $t$ submatrices:
\[\textbf{B}=\begin{pmatrix}
    \textbf{B}_1&\cdots&\textbf{B}_t
\end{pmatrix},\ {\bf{\Lambda}}^T=\begin{pmatrix}
    {\bf{\Lambda}}_1^T&\cdots&{\bf{\Lambda}}_t^T
\end{pmatrix},\]
where $\textbf{B}_i\in\mathbb{F}_q^{(T_u+m)\times K},{\bf{\Lambda}}_i\in\mathbb{F}_q^{K\times K}$.
Exploiting the cyclic structure of each subnetwork (for $i \in [t]$), we define each ${\bf{\Lambda}}_i$ as follows:
\[{\bf{\Lambda}}_i=\begin{pmatrix}
    a_1&a_2&\cdots&a_K\\
    a_K&a_1&\cdots&a_{K-1}\\
    \vdots&\vdots&\ddots&\vdots\\
    a_2&a_3&\cdots&a_1
\end{pmatrix},\]
and ${\bf{\Lambda}}_i={\bf{\Lambda}}_{i'}$ for any $i,i'\in[t]$. Moreover, we choose $a_1$ to be sufficiently large relative to the magnitudes of the remaining entries $a_i$.

Let us determine the choice of $\textbf{B}$. We define $\textbf{B}=(b_{ij})$ for $i \in [T_u+m]$ and $j \in [N]$, where $b_{ij}=\frac{1}{\alpha_i+\beta_j}$. We set $\alpha_i=i$ and define 
\begin{equation}
    \beta_j=\begin{cases}
        c_j, &1\leq j\leq K,\\
        c_jw, &H+1\leq j\leq 2K,\\
         &\vdots\\
        c_jw^{t-1}, &(t-1)K+1\leq j\leq tK,
    \end{cases}
\end{equation}
where $w$ is a primitive $t$-th root of unity in $\mathbb{F}_q$. With this construction, $\textbf{B}$ is a Cauchy matrix, which is known to be an MDS matrix.

Now, with $\textbf{B}$ and $\bf{\Lambda}$ defined, we can determine $\textbf{D}$ using the condition from (\ref{eq:BlambdaD=0}). Specifically, we have
\[\textbf{B}\cdot{\bf{\Lambda}}\cdot \textbf{D}^T=\begin{pmatrix}
    \textbf{B}_1&\cdots&\textbf{B}_t
\end{pmatrix}\begin{pmatrix}
    {\bf{\Lambda}}_1\\ \vdots\\ {\bf{\Lambda}}_t
\end{pmatrix}\textbf{D}^T=\left(\sum_{k=1}^{t}\textbf{B}_k{\bf{\Lambda}}_k\right)\textbf{D}^T=\left(\sum_{k=1}^{t}\textbf{B}_k\right){\bf{\Lambda}}_1\textbf{D}^T=\Zero_{(T_u+m)\times n}.\]
In the above step, we have used the fact that ${\bf{\Lambda}}_k = {\bf{\Lambda}}_1$ for all $k \in [t]$.

Considering the structure of $\textbf{B}$, the sum $\sum_{k=1}^{t}\textbf{B}_k$ can be simplified. For the element at row $i$ and column $j$ of $\sum_{k=1}^{t}\textbf{B}_k$, we observe that
\begin{subequations}
    \begin{align}
        \sum_{k=1}^{t}\frac{1}{\alpha_i+\beta_{j+(k-1)K}}&=\frac{\sum_{k=1}^{t}\prod_{k'\in[t]\backslash\{k\}}(\alpha_i+\beta_{j+(k-1)K})}{\prod_{k\in[t]}(\alpha_i+\beta_{j+(k-1)H})}\\
        &=\frac{t\alpha_i^{t-1}}{\alpha_i^t+\beta_j\cdots \beta_{j+(t-1)K}}\\
        &=\frac{t\alpha_i^{t-1}}{\alpha_i^t+c_j^t}.
\end{align}
\end{subequations}
This derivation shows that $\sum_{k=1}^{t}\textbf{B}_k$ retains the form of a Cauchy matrix and is therefore also an MDS matrix.

We now introduce an MDS matrix $\textbf{Q} \in \mathbb{F}_q^{K \times n}$ such that $(\sum_{k=1}^{t}\textbf{B}_k) \cdot \textbf{Q} = \mathbf{0}_{(T_u+m)\times n}$. This construction is feasible because we can augment the matrix $(\sum_{k=1}^t \textbf{B}_k)$ by adding $K - (m + T_u)$ rows to form an invertible matrix.  Specifically, we define:
\[\bar{\textbf{B}}\eqdef\begin{pmatrix}
    \sum_{k=1}^{t}\textbf{B}_k\\
    \bb_{T_u+m+1}\\
    \vdots\\
    \bb_{K}
\end{pmatrix}=\begin{pmatrix}
    \bb_1\\ \vdots\\ \bb_{T_u+m}\\\bb_{T_u+m+1}\\
    \vdots\\
    \bb_{K}
\end{pmatrix}.\]
Since $\bar{\textbf{B}}\bar{\textbf{B}}^{-1}=\textbf{I}_{K\times K}$, we set $\textbf{Q}$ as the matrix formed by selecting the last $n$ columns of $\bar{\textbf{B}}^{-1}$. Consequently, $(\sum_{k=1}^{t}\textbf{B}_k)\textbf{Q}=\Zero_{(T_u+m)\times n}$, and it follows that $\textbf{Q}$ is also an MDS matrix.

Let ${\bf{\Lambda}}_1\textbf{D}^T=\textbf{Q}$, then we have
\[a_1\bd_1+a_2\bd_2+\cdots+a_K\bd_K=\textbf{q}_1,\]
where $\bd_i$ and $\textbf{q}_i$ represent the $i$-th rows of $\textbf{D}^T$ and $\textbf{Q}$, respectively. Given the condition that $a_1$ is sufficiently large compared to the other $a_i$ values, we can approximate
\[\bd_i\approx\textbf{q}_i/a_1.\]
Therefore, the established MDS property of $\textbf{Q}$ is effectively transferred to $\textbf{D}^T$ through this construction.

Finally, since all three conditions (P1),(P2) and (P3) are satisfied, the decodability and $(0, 1, T_u)$-security of this scheme are guaranteed.

\subsection{Scheme for Theorem~\ref{thm:caseT=N-3}}
In this subsection, we present the construction of a $(0,1,N-3)$-secure scheme for a cyclic network where $N=K$ and $n=m=2$. This scheme achieves the key rate pair $(R_Z,R_{Z_\Sigma})=(\frac{1}{2},\frac{N-1}{2})$. Analogous to the approach detailed in Subsection~\ref{subsec:scheme(1/n)}, the primary task involves carefully selecting the matrices $\textbf{B}\in\mathbb{F}_q^{(N-1)\times N}$, ${\bf{\Lambda}}\in\mathbb{F}_q^{N\times N}$, and $\textbf{D}\in\mathbb{F}_q^{2\times N}$ such that they collectively satisfy conditions (P1), (P2), and (P3).

For a prime $q\geq N+2$, we select the matrix $\textbf{D}$ as
\[\textbf{D}=\begin{pmatrix}
    1&1&\cdots&1\\
    1&2&\cdots&N
\end{pmatrix}.\]
This construction ensures that $\textbf{D}$ is an MDS matrix. Given that $n=m=2$, the matrix $\bf{\Lambda}$ must contain exactly two non-zero elements in each row and each column. Since the network is cyclic, we can assume that for $i\in[N]$, $\cH_i=\{\Mod(i,N),\Mod(i+1,N)\}$.  For $i\in[N-1]$, define $\lambda_i=\frac{i-N-1}{N-i}$, and let $\lambda_N=\frac{1}{N}$. We select $\bf{\Lambda}$ as
\[{\bf{\Lambda}}=\begin{pmatrix}
    1&\lambda_1&0&\cdots&0\\
    0&1&\lambda_2&\cdots&0\\
    \vdots&\vdots&\ddots&\ddots&\vdots\\
    \lambda_N&0&0&\cdots&1
\end{pmatrix}.\]
With this choice, condition (P2) is naturally satisfied.
Next, we define the matrix $\textbf{B}$ as:
\[\textbf{B}=\begin{pmatrix}
    \textbf{I}_{N-1}&\bb_N
\end{pmatrix}=\begin{pmatrix}
    1&0&\cdots&0&-\frac{1+\lambda_1}{1+\lambda_N}\\
    0&1&\cdots&0&-\frac{1+\lambda_2}{1+\lambda_N}\\
    \vdots&\vdots&\ddots&\vdots&\vdots\\
    0&0&\cdots&1&-\frac{1+\lambda_{N-1}}{1+\lambda_N}
\end{pmatrix}.\]
For $q\geq N+2$, we have $-\frac{1+\lambda_i}{1+\lambda_N}=\frac{N}{(N+1)(N-i)} \neq 0$ for all $i \in [N-1]$. This guarantees that $\textbf{B}$ is an MDS matrix, thereby satisfying condition (P1).

Finally, we verify condition (P3):
\begin{subequations}
    \begin{align}
        \textbf{B}\cdot{\bf{\Lambda}}\cdot \textbf{D}^T=&\begin{pmatrix}
    \textbf{I}_{N-1}&\bb_N
\end{pmatrix}\cdot\begin{pmatrix}
    1+\lambda_1&1+2\lambda_1\\
    1+\lambda_2&2+3\lambda_2\\
    \vdots&\vdots\\
    1+\lambda_{N-1}&N-1+N\lambda_{N-1}\\
    1+\lambda_N&N+\lambda_N
\end{pmatrix}\\
=&\begin{pmatrix}
    \textbf{I}_{N-1}&\bb_N
\end{pmatrix}\cdot\begin{pmatrix}
    1+\lambda_1&(N+1)(1+\lambda_1)\\
    1+\lambda_2&(N+1)(1+\lambda_2)\\
    \vdots&\vdots\\
    1+\lambda_N&(N+1)(1+\lambda_N)
\end{pmatrix}\\
=&\Zero_{N\times2}.
    \end{align}
\end{subequations}
As demonstrated, condition (P3) is also satisfied. Consequently, the scheme constructed using these matrices $\textbf{B}$, $\bf{\Lambda}$, and $\textbf{D}$ is a $(0,1,N-3)$-secure scheme, which completes the proof of Theorem~\ref{thm:caseT=N-3}.

\section{Further anaylsis of key rates in Example~\ref{eg:example_3,3,2,2_}}\label{sec:example}
In this section, we first introduce an illustrative example to demonstrate that the lower bound for key rates in Theorem~\ref{thm:lowerboundR_Z} is not always tight. Furthermore, we extend the lower bound for this example to general cyclic networks with $n=m=2$.
\subsection{An illustrative example}
In Example~\ref{eg:example_3,3,2,2_}, we present a $(0,1,1)$-secure scheme with optimal communication load with $(R_Z,R_{Z_\Sigma})=(1,2)$ for network $\cN_{3,3,2}$. For such a network, from Theorem~\ref{thm:lowerboundR_Z}, we have $R_Z\geq\frac{1}{2}$. 
In this subsection, we will prove that for the HSA problem in Example~\ref{eg:example_3,3,2,2_}, if there exists a scheme attaining optimal communication load, then the scheme must have key rate $R_Z\geq 1$ and $R_{Z_\Sigma}\geq 2$, which implies the scheme presented in Example~\ref{eg:example_3,3,2,2_} is optimal.

The key improvement in establishing the lower bound stems from the observation that the value of $Z_{i,j}$ is implicitly determined by the set $\{Z_{i'}: i' \neq i\}$. This relationship is formally captured in the following lemma.
\begin{lem}\label{lem:Z_dependence}
If the communication load is optimized, then
    \begin{equation}
        H(Z_{i,j}|Z_{[3]\backslash\{i\}})=0.
    \end{equation}
\end{lem}
\begin{pf}
    Take $Z_{3,3}$ as an example.
    Recall that $Z_{i,j}, W_{i,j}$ are the subsets of symbols of $Z_i$ and $W_i$ contained in $X_{i,j}$ respectively. 
    From Proposition~\ref{prop:zijwij}, we know that
    \begin{equation}\label{eq:lemma3_pre0}
        H(Z_{i,j}|X_{i,j},W_{i,j})=0. 
    \end{equation}
    From the decodability, we have
    \begin{subequations}\label{eq:lemma3_pre1}
        \begin{align}
            0&=H(W_1+W_2+W_3|Y_{[3]})\\
            &\geq H(W_1+W_2+W_3|Y_{[3]},\{X_{i,j}\}_{i\in[3],j\in\cH_i})\\
            &= H(W_1+W_2+W_3|\{X_{i,j}\}_{i\in[3],j\in\cH_i})\\
            &\geq H(W_1+W_2+W_3|\{X_{i,j}\}_{i\in[3],j\in\cH_i},W_{[2]},Z_{[2]})\\
            &= H(W_1+W_2+W_3|X_{3,1},X_{3,3},W_{[2]},Z_{[2]})\\
            &=H(W_3|X_{3,1},X_{3,3},W_{[2]},Z_{[2]})\\
            &=H(W_3|W_{[2]},Z_{[2]})-I(W_3;X_{3,1},X_{3,3}|W_{[2]},Z_{[2]})\\
            &=H(W_3)-I(W_3;X_{3,1},X_{3,3}|W_{[2]},Z_{[2]}),
        \end{align}
    \end{subequations}
    (\ref{eq:lemma3_pre1}b)-(\ref{eq:lemma3_pre1}e) hold due to the principle that additional conditions cannot increase entropy, coupled with the fact that \(Y_{[3]}\) is a function of \(\{X_{i,j}\}\), and \(\{X_{1,j}, X_{2,j}\}\) are functions of \(W_1, W_2, Z_1, Z_2\). From (\ref{eq:lemma3_pre1}), we have
    \begin{equation}\label{eq:lemma3_pre2}
        I(W_3;X_{3,1},X_{3,3}|W_{[2]},Z_{[2]})=H(W_3).
    \end{equation}
    If the communication load is optimized, from Lemma~\ref{lem:xijyj}, we have \[H(X_{3,1},X_{3,3}|W_1,W_2,Z_1,Z_2)=L,\] then we have
    \begin{subequations}
        \begin{align}
            L&=H(W_3)\overset{(\ref{eq:lemma3_pre2})}{=} I(W_3;X_{3,1},X_{3,3}|W_1,W_2,Z_1,Z_2)\\
            &=H(X_{3,1},X_{3,3}|W_1,W_2,Z_1,Z_2)-H(X_{3,1},X_{3,3}|W_{[3]},Z_1,Z_2)\\
            &= L-H(X_{3,1},X_{3,3}|W_{[3]},Z_1,Z_2),
        \end{align}
    \end{subequations}
    which implies   \begin{equation}\label{eq:lemma3_pre5}
        H(X_{3,1},X_{3,3}|W_{[3]},Z_1,Z_2)=0.
    \end{equation}
    Combining (\ref{eq:lemma3_pre0}) and (\ref{eq:lemma3_pre5}), we have
    \begin{subequations}\label{eq:lemma3}
        \begin{align}
            H(Z_{3,3}|Z_1,Z_2)&=H(Z_{3,3}|W_{[3]},Z_1,Z_2)\\
            &\leq H(Z_{3,3},W_{3,3},X_{3,3}|W_{[3]},Z_1,Z_2)\\
            &\overset{(\ref{eq:lemma3_pre0})}{=}H(W_{3,3},X_{3,3}|W_{[3]},Z_1,Z_2)\\
            &=H(X_{3,3}|W_{[3]},Z_1,Z_2)\\
            &\leq H(X_{3,3},X_{3,1}|W_{[3]},Z_1,Z_2)\overset{(\ref{eq:lemma3_pre5})}{=}0,
        \end{align}
    \end{subequations}
    where (\ref{eq:lemma3}a) is due to the independence between $\{W_i\}$ and $\{Z_j\}$.
\end{pf}

Using this lemma, we obtain the following theorem.
\begin{thm}~\label{thm:Rz}
If the communication load is optimized, then
    \begin{equation}
        H(Z_1)+H(Z_2)+H(Z_3)\geq 3L.
    \end{equation}
\end{thm}
\begin{pf}
    \begin{subequations}\label{eq:exampleRz}
        \begin{align}
            H(Z_2)&\geq H(Z_2|Z_1)\\
            &=H(Z_2|Z_1)+H(Z_{3,3}|Z_1,Z_2)\\
            &=H(Z_2,Z_{3,3}|Z_1)\\
            &\geq H(Z_{2,3},Z_{3,3}|Z_1),
        \end{align}
    \end{subequations}
    where $(\ref{eq:exampleRz}b)$ is due to Lemma~\ref{lem:Z_dependence}. Similarly, we have
    \begin{align}
        H(Z_1)\geq H(Z_{3,1},Z_{1,1}|Z_2),\\
        H(Z_3)\geq H(Z_{1,2},Z_{2,2}|Z_3).
    \end{align}
    Therefore, from Lemma~\ref{lem:ZijtoXij}, we have
    \begin{equation*}
        H(Z_1)+H(Z_2)+H(Z_3)\geq 3L.
    \end{equation*}
\end{pf}

\emph{Proof of $R_Z\geq1$}:
    From Theorem~\ref{thm:Rz}, 
    \begin{equation*}
        3R_ZL\geq H(Z_1)+H(Z_2)+H(Z_3)\geq3L.
    \end{equation*}
    
\emph{Proof of $R_{Z_\Sigma}\geq 2$}:
    Let $\mathcal{T}_j$ be the user colluded with the relay $j$. Let $\mathcal{T}_1=\{3\},\mathcal{T}_2=\{1\},\mathcal{T}_3=\{2\}$, then
    \begin{align*}
        3H(Z_{\Sigma})&\geq H(Z_{1,1},Z_{2,1},Z_3)+H(Z_{2,2},Z_{3,1},Z_1)+H(Z_{1,3},Z_{3,3},Z_2)\\
        &\geq \sum_{i\in[3]}Z_i+H(Z_{1,1},Z_{2,1}|Z_3)+H(Z_{2,2},Z_{3,2}|Z_1)+H(Z_{1,3},Z_{3,3}|Z_2)\\
        &\geq 6L.
    \end{align*}

\subsection{Proof of Theorem~\ref{thm:casen=m=2}}
In this subsection, we extend the proof in Section~\ref{sec:example} to any cyclic network with $n=m=2$. We first extend Lemma~\ref{lem:Z_dependence} to general $N,K$.
\begin{lem}
    If the communication load is optimal, then 
    \[H(Z_{i,j}|Z_{[N]\backslash\{i\}})=0.\]
\end{lem}
\begin{pf}
    First, we note that (\ref{eq:lemma3_pre0}) holds for any network, since it is only dependent on the definition of $X_{i,j},W_{i,j}$ and $Z_{i,j}$. Next, from the decodability, we have
    \begin{subequations}\label{eq:genlemma3_pre1}
        \begin{align}
            0&=H(\sum_{i\in[N]}W_i|Y_{[K]})\\
            &\geq H(\sum_{i\in[N]}W_i|Y_{[K]},\{X_{i,j}\}_{i\in[N],j\in\cH_i})\\
            &= H(\sum_{i\in[N]}W_i|\{X_{i,j}\}_{i\in[N],j\in\cH_i})\\
            &\geq H(\sum_{i\in[N]}W_i|\{X_{i,j}\}_{i\in[N],j\in\cH_i},W_{[N]\backslash\{i\}},Z_{[N]\backslash\{i\}})\\
            &= H(\sum_{i\in[N]}W_i|\{X_{i,j}\}_{j\in\cH_i},W_{[N]\backslash\{i\}},Z_{[N]\backslash\{i\}})\\
            &=H(W_i|\{X_{i,j}\}_{j\in\cH_i},W_{[N]\backslash\{i\}},Z_{[N]\backslash\{i\}})\\
            &=H(W_i|W_{[N]\backslash\{i\}},Z_{[N]\backslash\{i\}})-I(W_i;\{X_{i,j}\}_{j\in\cH_i}|W_{[N]\backslash\{i\}},Z_{[N]\backslash\{i\}})\\
            &=H(W_i)-I(W_i;\{X_{i,j}\}_{j\in\cH_i}|W_{[N]\backslash\{i\}},Z_{[N]\backslash\{i\}}),
        \end{align}
    \end{subequations}
    (\ref{eq:genlemma3_pre1}b)-(\ref{eq:genlemma3_pre1}e) hold due to the principle that additional conditions cannot increase entropy, coupled with the fact that \(Y_{[K]}\) is a function of \(\{X_{i,j}\}\), and \(\{X_{i',j}\}_{i'\neq i,j\in\cH_{i'}}\) are functions of \(W_{[N]\backslash\{i\}}, Z_{[N]\backslash\{i\}}\). From (\ref{eq:genlemma3_pre1}), we have
    \begin{equation}\label{eq:genlemma3_pre2}
        I(W_i;\{X_{i,j}\}_{j\in\cH_i}|W_{[N]\backslash\{i\}},Z_{[N]\backslash\{i\}})=H(W_i).
    \end{equation}
    If the communication load is optimized, from Lemma~\ref{lem:xijyj}, we have \[H(\{X_{i,j}\}_{j\in\cH_i}|W_{[N]\backslash\{i\}},Z_{[N]\backslash\{i\}})=L.\] Then,
    \begin{subequations}
        \begin{align}
            L&=H(W_i)\overset{(\ref{eq:genlemma3_pre2})}{=}  I(W_i;\{X_{i,j}\}_{j\in\cH_i}|W_{[N]\backslash\{i\}},Z_{[N]\backslash\{i\}})\\
            &=H(\{X_{i,j}\}_{j\in\cH_i}|W_{[N]\backslash\{i\}},Z_{[N]\backslash\{i\}})-H(\{X_{i,j}\}_{j\in\cH_i}|W_{[N]},Z_{[N]\backslash\{i\}})\\
            &= L-H(\{X_{i,j}\}_{j\in\cH_i}|W_{[N]},Z_{[N]\backslash\{i\}}),
        \end{align}
    \end{subequations}
    which implies   \begin{equation}\label{eq:genlemma3_pre5}
        H(\{X_{i,j}\}_{j\in\cH_i}|W_{[N]},Z_{[N]\backslash\{i\}})=0.
    \end{equation}
    Combining (\ref{eq:lemma3_pre0}) and (\ref{eq:genlemma3_pre5}), we have
    \begin{subequations}\label{eq:genlemma3}
        \begin{align}
            H(Z_{i,j}|Z_{[N]\backslash\{i\}})&=H(Z_{i,j}|W_{[N]},Z_{[N]\backslash\{i\}})\\
            &\leq H(Z_{i,j},W_{i,j},X_{i,j}|W_{[N]},Z_{[N]\backslash\{i\}})\\
            &\overset{(\ref{eq:lemma3_pre0})}{=}H(W_{i,j},X_{i,j}|W_{[N]},Z_{[N]\backslash\{i\}})\\
            &=H(X_{i,j}|W_{[N]},Z_{[N]\backslash\{i\}})\\
            &\leq H(\{X_{i,j}\}_{j\in\cH_i}|W_{[N]},Z_{[N]\backslash\{i\}})\overset{(\ref{eq:genlemma3_pre5})}{=}0.
        \end{align}
    \end{subequations}
    Equation (\ref{eq:genlemma3}a) is due to the independence between $\{W_i\}$ and $\{Z_j\}$. The proof is finished.
\end{pf}

Now, we are ready to show that for a cyclic network with $n=2$, if $T_h=1,T_u=N-2$, then
$H(Z_i)\geq L$ and $H(Z_{\Sigma})\geq N-1$.
We first define some notations. For any relay $j\in[K]$, let $\bar{\cU}_j\eqdef[N]\backslash\cU_j$.
Since the network is cyclic, we can obtain that for any $j\in[K]$, 
\begin{align}
    &\cU_j=\{\Mod(j,N),\Mod(j+1,N)\},\\
    &\bar{\cU}_j=\{\Mod(j+2,N),\Mod(j+N-1,N)\}.
\end{align}
For simplicity, we omit the modular symbol, and denote 
\begin{align}
    &\cU_j=\{j,j+1\},\\
    &\bar{\cU}_j=\{j+2,\cdots,j+N-1\}.
\end{align}

\emph{Proof of $R_Z\geq1$}:  For any $j\in[K]$,
\begin{subequations}\label{eq:genexampleRZ}
    \begin{align}
        H(Z_j)\geq& H(Z_j|Z_{\bar{\cU}_j})\\
        =&H(Z_j|Z_{\bar{\cU}_j})+H(Z_{j+1,j}|Z_{\bar{\cU}_j},Z_{j})\\
        =&H(Z_{j},Z_{j+1,j}|Z_{\bar{\cU}_j})\\
        \geq& H(Z_{j,j},Z_{j+1,j}|Z_{\bar{\cU}_j}),
    \end{align}
\end{subequations}
where $(\ref{eq:genexampleRZ}b)$ is due to Lemma~\ref{lem:Z_dependence}. Since the network is cyclic, we have
\begin{subequations}\label{eq:pfofRZ>1}
    \begin{align}
        \sum_{j\in[K]}H(Z_j)\geq&\sum_{j\in[K]}H(\{Z_{i,j}\}_{i\in\cU_j}|Z_{\bar{\cU}_j})\\
        \geq&\sum_{j\in[K]}\sum_{i\in\cU_j}H(X_{i,j}|\{W_{i'},Z_{i'}\}_{i'\neq i})\\
        \geq&\sum_{i\in[N]}H(\{X_{i,j}\}_{j\in\cH_i}|\{W_{i'},Z_{i'}\}_{i'\neq i})\\
        \geq & NL,
    \end{align}
\end{subequations}
where (\ref{eq:pfofRZ>1}b) is due to Lemma~\ref{lem:ZijtoXij} and (\ref{eq:pfofRZ>1}d) is due to Lemma~\ref{lem:xijyj}.
Thus, we know that $\max_{j}H(Z_j)\geq L$, i.e., $R_Z\geq1$.

\emph{Proof of $R_{Z_\Sigma}\geq N-1$}:
For any $j\in[K]$
\begin{subequations}\label{eq:pfofRZsum1}
    \begin{align}
        H(Z_{\bar{\cU}_j})=&H(Z_{j+2},\cdots,Z_{j+N-1})\\
        =&H(Z_{j+2})+H(Z_{j+3}|Z_{j+2})+\cdots+H(Z_{j+N-1}|Z_{j+2},\cdots,Z_{j+N-2})\\
        \geq &H(Z_{j+2}|Z_{\bar{\cU}_{j+2}})+H(Z_{j+3}|Z_{\bar{\cU}_{j+3}})+\cdots+H(Z_{j+N-1}|Z_{\bar{\cU}_{j+N-1}})\\
        =&\sum_{t=j+2}^{j+N-1}H(Z_t|Z_{\bar{\cU}_t}),
    \end{align}
\end{subequations}
where (\ref{eq:pfofRZsum1}b) is due to the chain rule of joint entropy, and (\ref{eq:pfofRZsum1}c) is because the condition part is contained in $Z_{\bar{\cU}_t}$. 
Therefore, we can obtain
\begin{subequations}\label{eq:pfofRZsum2}
    \begin{align}
        KH(Z_{\Sigma})&\geq\sum_{j\in[K]}H(Z_{\bar{\cU}_j},Z_{j,j},Z_{j+1,j})\\
        &=\sum_{j\in[K]}H(Z_{\bar{\cU}_j})+\sum_{j\in[K]}H(Z_{j,j},Z_{j+1,j}|Z_{\bar{\cU}_j})\\
        &\overset{(\ref{eq:pfofRZsum1})}{\geq} \sum_{j\in[K]}\sum_{t=j+2}^{j+N-1}H(Z_t|Z_{\bar{\cU}_t})+\sum_{j\in[K]}H(Z_{j,j},Z_{j+1,j}|Z_{\bar{\cU}_j})\\
        &=(N-2)\sum_{j\in[K]}H(Z_j|Z_{\bar{\cU}_j})+\sum_{j\in[K]}H(Z_{j,j},Z_{j+1,j}|Z_{\bar{\cU}_j})\\
        &\overset{(\ref{eq:genexampleRZ})(\ref{eq:pfofRZ>1})}{\geq} (N-2)NL+\sum_{j\in[K]}H(Z_{j,j},Z_{j+1,j}|Z_{\bar{\cU}_j})\\
        &\overset{(\ref{eq:pfofRZ>1})}{\geq}(N-2)NL+NL=(N-1)NL,
    \end{align}
\end{subequations}
where (\ref{eq:pfofRZsum2}a) is from the definition of $Z_\Sigma$ and (\ref{eq:pfofRZsum2}d) is because each term $H(Z_j|Z_{\bar{\cU}_j})$ is counted by $N-2$ times. Finally, since $K=N$, we have $H(Z_{\Sigma})\geq (N-1)L$, i.e., $R_{Z_\Sigma}\geq N-1$. Thus, we finish the proof of Theorem~\ref{thm:casen=m=2}.

\section{Conclusion}\label{sec:conclusion}
This paper addressed the HSA problem for general user-relay association under scenarios involving collusion between relays and users. We derived a lower bound on the communication load and established that no $(0,T_h,T_u)$-secure scheme achieving optimal communication load exists if $T_h \geq K-n+1$ or $T_u \geq n(\cN_{N,K,n},T_h)$. Furthermore, for the regime where $0 < T_h \leq K-n$ and $T_u \leq n(\cN_{N,K,n},T_h)-1$, we proposed a scheme based on network function computation that attains optimal communication load. Consequently, the main challenge for this problem lies in reducing the required key size. To address this, we first presented lower bounds for $R_Z$ and $R_{Z_\Sigma}$, and subsequently developed a scheme for multiple cyclic networks that simultaneously achieves both optimal communication load and optimal key size. 
However, the derived lower bound for $R_Z$ and $R_{Z_{\Sigma}}$ is neither universally applicable across all parameter regimes nor consistently tight. Moreover, the scheme achieving optimal key size requires small values of $T_u$ and $T_h$. We illustrated these limitations through an example where $R_Z$ and $R_{Z_\Sigma}$ strictly exceed the lower bound while remaining achievable.
For cyclic networks with $n=m=2$, we obtained a tighter lower bound and provided a corresponding scheme attaining optimal key rates.

Several future directions can be considered: 1) The behavior of $R_Z$ and $R_{Z_\Sigma}$ under large $T_h$ and $T_u$ for a general network constitutes; 2) Another significant scenario involves $(1,T_h,T_u)$-security, representing a more general security constraint. Under this stronger requirement, colluding nodes inevitably learn the sum of source messages but must gain no further information about individual messages.

\bibliographystyle{IEEEtranS}
\bibliography{reference}

@article{2025HSAcyclic,
  title={Fundamental Limits of Hierarchical Secure Aggregation with Cyclic User Association},
  author={Zhang, Xiang and Li, Zhou and Wan, Kai and Sun, Hua and Ji, Mingyue and Caire, Giuseppe},
  journal={arXiv preprint arXiv:2503.04564},
  year={2025}
}

@inproceedings{2017bonawitzpractical,
  title={Practical secure aggregation for privacy-preserving machine learning},
  author={Bonawitz, Keith and Ivanov, Vladimir and Kreuter, Ben and Marcedone, Antonio and McMahan, H Brendan and Patel, Sarvar and Ramage, Daniel and Segal, Aaron and Seth, Karn},
  booktitle={proceedings of the 2017 ACM SIGSAC Conference on Computer and Communications Security},
  pages={1175--1191},
  year={2017}
}

@ARTICLE{2022ZhaoSun,
  author={Zhao, Yizhou and Sun, Hua},
  journal={IEEE Transactions on Information Theory}, 
  title={Information Theoretic Secure Aggregation With User Dropouts}, 
  year={2022},
  volume={68},
  number={11},
  pages={7471-7484},
  doi={10.1109/TIT.2022.3192874}}

@inproceedings{2017mcmahancommunication,
  title={Communication-efficient learning of deep networks from decentralized data},
  author={McMahan, Brendan and Moore, Eider and Ramage, Daniel and Hampson, Seth and y Arcas, Blaise Aguera},
  booktitle={Artificial intelligence and statistics},
  pages={1273--1282},
  year={2017},
  organization={PMLR}
}

@article{2016konevcnyfederated,
  title={Federated learning: Strategies for improving communication efficiency},
  author={Kone{\v{c}}n{\`y}, Jakub and McMahan, H Brendan and Yu, Felix X and Richt{\'a}rik, Peter and Suresh, Ananda Theertha and Bacon, Dave},
  journal={arXiv preprint arXiv:1610.05492},
  year={2016}
}

@article{2021kairouzadvances,
  title={Advances and open problems in federated learning},
  author={Kairouz, Peter and McMahan, H Brendan and Avent, Brendan and Bellet, Aur{\'e}lien and Bennis, Mehdi and Bhagoji, Arjun Nitin and Bonawitz, Kallista and Charles, Zachary and Cormode, Graham and Cummings, Rachel and others},
  journal={Foundations and trends in machine learning},
  volume={14},
  number={1--2},
  pages={1--210},
  year={2021},
  publisher={Now Publishers, Inc.}
}

@ARTICLE{2020WeiLiDing,
  author={Wei, Kang and Li, Jun and Ding, Ming and Ma, Chuan and Yang, Howard H. and Farokhi, Farhad and Jin, Shi and Quek, Tony Q. S. and Vincent Poor, H.},
  journal={IEEE Transactions on Information Forensics and Security}, 
  title={Federated Learning With Differential Privacy: Algorithms and Performance Analysis}, 
  year={2020},
  volume={15},
  number={},
  pages={3454-3469},
  doi={10.1109/TIFS.2020.2988575}}

@ARTICLE{2020HuGuoLi,
  author={Hu, Rui and Guo, Yuanxiong and Li, Hongning and Pei, Qingqi and Gong, Yanmin},
  journal={IEEE Internet of Things Journal}, 
  title={Personalized Federated Learning With Differential Privacy}, 
  year={2020},
  volume={7},
  number={10},
  pages={9530-9539},
  doi={10.1109/JIOT.2020.2991416}}

@ARTICLE{2021ZhaoZhaoYang,
  author={Zhao, Yang and Zhao, Jun and Yang, Mengmeng and Wang, Teng and Wang, Ning and Lyu, Lingjuan and Niyato, Dusit and Lam, Kwok-Yan},
  journal={IEEE Internet of Things Journal}, 
  title={Local Differential Privacy-Based Federated Learning for Internet of Things}, 
  year={2021},
  volume={8},
  number={11},
  pages={8836-8853},
  doi={10.1109/JIOT.2020.3037194}}

@article{2021andrewdifferentially,
  title={Differentially private learning with adaptive clipping},
  author={Andrew, Galen and Thakkar, Om and McMahan, Brendan and Ramaswamy, Swaroop},
  journal={Advances in Neural Information Processing Systems},
  volume={34},
  pages={17455--17466},
  year={2021}
}

@ARTICLE{2024YeminiSaha,
  author={Yemini, Michal and Saha, Rajarshi and Ozfatura, Emre and Gündüz, Deniz and Goldsmith, Andrea J.},
  journal={IEEE Transactions on Wireless Communications}, 
  title={Robust Semi-Decentralized Federated Learning via Collaborative Relaying}, 
  year={2024},
  volume={23},
  number={7},
  pages={7520-7536},
  doi={10.1109/TWC.2023.3342095}}

@article{2024lindifferential,
  title={Differential privacy in hierarchical federated learning: A formal analysis and evaluation},
  author={Lin, F Po-Chen and Brinton, C},
  journal={arXiv preprint	arXiv:2401.11592},
  year={2024}
}

@ARTICLE{2021SoGuler,
  author={So, Jinhyun and Güler, Başak and Avestimehr, A. Salman},
  journal={IEEE Journal on Selected Areas in Information Theory}, 
  title={Turbo-Aggregate: Breaking the Quadratic Aggregation Barrier in Secure Federated Learning}, 
  year={2021},
  volume={2},
  number={1},
  pages={479-489},
  doi={10.1109/JSAIT.2021.3054610}}

@article{2020kadhefastsecagg,
  title={Fastsecagg: Scalable secure aggregation for privacy-preserving federated learning},
  author={Kadhe, Swanand and Rajaraman, Nived and Koyluoglu, O Ozan and Ramchandran, Kannan},
  journal={arXiv preprint arXiv:2009.11248},
  year={2020}
}

@ARTICLE{2022ElkordyAvestimehr,
  author={Elkordy, Ahmed Roushdy and Avestimehr, A. Salman},
  journal={IEEE Transactions on Communications}, 
  title={HeteroSAg: Secure Aggregation With Heterogeneous Quantization in Federated Learning}, 
  year={2022},
  volume={70},
  number={4},
  pages={2372-2386},
  doi={10.1109/TCOMM.2022.3151126}}

@ARTICLE{2023LiuGuoLam,
  author={Liu, Ziyao and Guo, Jiale and Lam, Kwok-Yan and Zhao, Jun},
  journal={IEEE Transactions on Information Forensics and Security}, 
  title={Efficient Dropout-Resilient Aggregation for Privacy-Preserving Machine Learning}, 
  year={2023},
  volume={18},
  number={},
  pages={1839-1854},
  doi={10.1109/TIFS.2022.3163592}}

@ARTICLE{2023JahaniMALi,
  author={Jahani-Nezhad, Tayyebeh and Maddah-Ali, Mohammad Ali and Li, Songze and Caire, Giuseppe},
  journal={IEEE Journal on Selected Areas in Communications}, 
  title={SwiftAgg+: Achieving Asymptotically Optimal Communication Loads in Secure Aggregation for Federated Learning}, 
  year={2023},
  volume={41},
  number={4},
  pages={977-989},
  doi={10.1109/JSAC.2023.3242702}}

@article{shamir1979share,
  title={How to share a secret},
  author={Shamir, Adi},
  journal={Communications of the ACM},
  volume={22},
  number={11},
  pages={612--613},
  year={1979},
  publisher={ACm New York, NY, USA}
}

@article{shannon1949communication,
  title={Communication theory of secrecy systems},
  author={Shannon, Claude E},
  journal={The Bell system technical journal},
  volume={28},
  number={4},
  pages={656--715},
  year={1949},
  publisher={Nokia Bell Labs}
}

@INPROCEEDINGS{2022JahaniMALi,
  author={Jahani-Nezhad, Tayyebeh and Maddah-Ali, Mohammad Ali and Li, Songze and Caire, Giuseppe},
  booktitle={2022 IEEE International Symposium on Information Theory (ISIT)}, 
  title={SwiftAgg: Communication-Efficient and Dropout-Resistant Secure Aggregation for Federated Learning with Worst-Case Security Guarantees}, 
  year={2022},
  volume={},
  number={},
  pages={103-108},
  doi={10.1109/ISIT50566.2022.9834750}}

@ARTICLE{2024Zhaosun,
  author={Zhao, Yizhou and Sun, Hua},
  journal={IEEE Transactions on Information Theory}, 
  title={Secure Summation: Capacity Region, Groupwise Key, and Feasibility}, 
  year={2024},
  volume={70},
  number={2},
  pages={1376-1387},
  doi={10.1109/TIT.2023.3342571}}

@ARTICLE{2024WanYaoSunJiCaire,
  author={Wan, Kai and Yao, Xin and Sun, Hua and Ji, Mingyue and Caire, Giuseppe},
  journal={IEEE Transactions on Information Theory}, 
  title={On the Information Theoretic Secure Aggregation With Uncoded Groupwise Keys}, 
  year={2024},
  volume={70},
  number={9},
  pages={6596-6619},
  doi={10.1109/TIT.2024.3422087}}

@ARTICLE{2024WanSunJiCaire,
  author={Wan, Kai and Sun, Hua and Ji, Mingyue and Mi, Tiebin and Caire, Giuseppe},
  journal={IEEE Transactions on Information Theory}, 
  title={The Capacity Region of Information Theoretic Secure Aggregation With Uncoded Groupwise Keys}, 
  year={2024},
  volume={70},
  number={10},
  pages={6932-6949},
  doi={10.1109/TIT.2024.3393740}}

@INPROCEEDINGS{2023ZhaoSunISIT,
  author={Zhao, Yizhou and Sun, Hua},
  booktitle={2023 IEEE International Symposium on Information Theory (ISIT)}, 
  title={The Optimal Rate of MDS Variable Generation}, 
  year={2023},
  volume={},
  number={},
  pages={832-837},
  doi={10.1109/ISIT54713.2023.10206914}}

@ARTICLE{2025ZhaoSunTIT,
  author={Zhao, Yizhou and Sun, Hua},
  journal={IEEE Transactions on Information Theory}, 
  title={MDS Variable Generation and Secure Summation With User Selection}, 
  year={2025},
  volume={71},
  number={4},
  pages={3129-3141},
  doi={10.1109/TIT.2025.3541551}}

@INPROCEEDINGS{2023LiZhaoSun,
  author={Li, Zhou and Zhao, Yizhou and Sun, Hua},
  booktitle={2023 IEEE International Symposium on Information Theory (ISIT)}, 
  title={Weakly Secure Summation with Colluding Users}, 
  year={2023},
  volume={},
  number={},
  pages={2398-2403},
  doi={10.1109/ISIT54713.2023.10206930}}

@article{2023sunsecure,
  title={Secure aggregation with an oblivious server},
  author={Sun, Hua},
  journal={arXiv preprint arXiv:2307.13474},
  year={2023}
}

@article{2025yuanSun,
  title={Vector linear secure aggregation},
  author={Yuan, Xihang and Sun, Hua},
  journal={arXiv preprint arXiv:2502.09817},
  year={2025}
}

@inproceedings{2021karakoccsecure,
  title={Secure aggregation against malicious users},
  author={Karako{\c{c}}, Ferhat and {\"O}nen, Melek and Bilgin, Zeki},
  booktitle={Proceedings of the 26th ACM symposium on access control models and technologies},
  pages={115--124},
  year={2021}
}

@article{2024zhangwansunwangoptimal,
  title={Optimal communication and key rate region for hierarchical secure aggregation with user collusion},
  author={Zhang, Xiang and Wan, Kai and Sun, Hua and Wang, Shiqiang and Ji, Mingyue and Caire, Giuseppe},
  journal={arXiv preprint arXiv:2410.14035},
  year={2024}
}

@INPROCEEDINGS{2023EggerHofmeister,
  author={Egger, Maximilian and Hofmeister, Christoph and Wachter-Zeh, Antonia and Bitar, Rawad},
  booktitle={2023 IEEE International Symposium on Information Theory (ISIT)}, 
  title={Private Aggregation in Wireless Federated Learning with Heterogeneous Clusters}, 
  year={2023},
  volume={},
  number={},
  pages={54-59},
  doi={10.1109/ISIT54713.2023.10206717}}

@ARTICLE{egger2023private,
  author={Egger, Maximilian and Hofmeister, Christoph and Wachter-Zeh, Antonia and Bitar, Rawad},
  journal={IEEE Transactions on Information Theory}, 
  title={Private Aggregation in Hierarchical Wireless Federated Learning With Partial and Full Collusion}, 
  year={2025},
  volume={71},
  number={11},
  pages={8977-8992},
  doi={10.1109/TIT.2025.3606645}}

@INPROCEEDINGS{2024ZhangWanSunWangITW,
  author={Zhang, Xiang and Wan, Kai and Sun, Hua and Wang, Shiqiang and Ji, Mingyue and Caire, Giuseppe},
  booktitle={2024 IEEE Information Theory Workshop (ITW)}, 
  title={Optimal Rate Region for Key Efficient Hierarchical Secure Aggregation with User Collusion}, 
  year={2024},
  volume={},
  number={},
  pages={573-578},
  doi={10.1109/ITW61385.2024.10806947}}

@article{2024Luchengkangliucapacity,
  title={Capacity of hierarchical secure coded gradient aggregation with straggling communication links},
  author={Lu, Qinyi and Cheng, Jiale and Kang, Wei and Liu, Nan},
  journal={arXiv preprint arXiv:2412.11496},
  year={2024}
}

@article{2025LiZhangLvFan,
  title={Collusion-Resilient Hierarchical Secure Aggregation with Heterogeneous Security Constraints},
  author={Li, Zhou and Zhang, Xiang and Lv, Jiawen and Fan, Jihao and Chen, Haiqiang and Caire, Giuseppe},
  journal={arXiv preprint arXiv:2507.14768},
  year={2025}
}

@ARTICLE{2011AFKZ,
  author={Appuswamy, Rathinakumar and Franceschetti, Massimo and Karamchandani, Nikhil and Zeger, Kenneth},
  journal={IEEE Transactions on Information Theory},
  title={Network Coding for Computing: Cut-Set Bounds},
  year={2011},
  volume={57},
  number={2},
  pages={1015-1030},
  doi={10.1109/TIT.2010.2095070}}

@ARTICLE{2018HTYG,
  author={Huang, Cupjin and Tan, Zihan and Yang, Shenghao and Guang, Xuan},
  journal={IEEE Transactions on Information Theory},
  title={Comments on Cut-Set Bounds on Network Function Computation},
  year={2018},
  volume={64},
  number={9},
  pages={6454-6459},
  doi={10.1109/TIT.2018.2827405}}

@ARTICLE{2019GYYL,
  author={Guang, Xuan and Yeung, Raymond W. and Yang, Shenghao and Li, Congduan},
  journal={IEEE Transactions on Information Theory},
  title={Improved Upper Bound on the Network Function Computing Capacity},
  year={2019},
  volume={65},
  number={6},
  pages={3790-3811},
  doi={10.1109/TIT.2019.2893107}}

@ARTICLE{2012RD,
  author={Rai, Brijesh Kumar and Dey, Bikash Kumar},
  journal={IEEE Transactions on Information Theory},
  title={On Network Coding for Sum-Networks},
  year={2012},
  volume={58},
  number={1},
  pages={50-63},
  doi={10.1109/TIT.2011.2169532}}

@ARTICLE{2014Appuswamy,
  author={Appuswamy, Rathinakumar and Franceschetti, Massimo},
  journal={IEEE Transactions on Information Theory},
  title={Computing Linear Functions by Linear Coding Over Networks},
  year={2014},
  volume={60},
  number={1},
  pages={422-431},
  doi={10.1109/TIT.2013.2283075}}

@INPROCEEDINGS{2002CaiYeung,
  author={Ning Cai and Yeung, R.W.},
  booktitle={Proceedings of the IEEE Information Theory Workshop},
  title={Network coding and error correction},
  year={2002},
  volume={},
  number={},
  pages={119-122},
  doi={10.1109/ITW.2002.1115432}}

@ARTICLE{2011Cai,
  author={Cai, Ning and Chan, Terence},
  journal={Proceedings of the IEEE}, 
  title={Theory of Secure Network Coding}, 
  year={2011},
  volume={99},
  number={3},
  pages={421-437},
  doi={10.1109/JPROC.2010.2094592}}

@article{2010CaiYeung,
  title={Secure network coding on a wiretap network},
  author={Cai, Ning and Yeung, Raymond W},
  journal={IEEE Transactions on Information Theory},
  volume={57},
  number={1},
  pages={424--435},
  year={2010},
  publisher={IEEE}
}

@article{2005Bhattad,
  title={Weakly secure network coding},
  author={Bhattad, Kapil and Narayanan, Krishna R and others},
  journal={NetCod, Apr},
  volume={104},
  pages={8--20},
  year={2005}
}

@article{2008Harada,
  title={Strongly secure linear network coding},
  author={Harada, Kunihiko and Yamamoto, Hirosuke},
  journal={IEICE transactions on fundamentals of electronics, communications and computer sciences},
  volume={91},
  number={10},
  pages={2720--2728},
  year={2008},
  publisher={The Institute of Electronics, Information and Communication Engineers}
}

@INPROCEEDINGS{2021GBY,
  author={Guang, Xuan and Bai, Yang and Yeung, Raymond W.},
  booktitle={2021 IEEE International Symposium on Information Theory (ISIT)}, 
  title={Secure Network Function Computation}, 
  year={2021},
  volume={},
  number={},
  pages={66-71},
  doi={10.1109/ISIT45174.2021.9518256}}

@ARTICLE{2024Guangsourcesecure,
  author={Guang, Xuan and Bai, Yang and Yeung, Raymond W.},
  journal={IEEE Transactions on Information Theory}, 
  title={Secure Network Function Computation for Linear Functions—Part I: Source Security}, 
  year={2024},
  volume={70},
  number={1},
  pages={676-697},
  doi={10.1109/TIT.2023.3328454}}

@article{2025BaiGuangFunctionsecure,
  title={Secure Network Function Computation for Linear Functions, Part II: Target-Function Security},
  author={Bai, Yang and Guang, Xuan and Yeung, Raymond W},
  journal={arXiv preprint arXiv:2504.17514},
  year={2025}
}

@ARTICLE{2021CDCsurvey,
  author={Ng, Jer Shyuan and Lim, Wei Yang Bryan and Luong, Nguyen Cong and Xiong, Zehui and Asheralieva, Alia and Niyato, Dusit and Leung, Cyril and Miao, Chunyan},
  journal={IEEE Communications Surveys \& Tutorials}, 
  title={A Comprehensive Survey on Coded Distributed Computing: Fundamentals, Challenges, and Networking Applications}, 
  year={2021},
  volume={23},
  number={3},
  pages={1800-1837},
  doi={10.1109/COMST.2021.3091684}}

@ARTICLE{2018Lisongze,
  author={Li, Songze and Maddah-Ali, Mohammad Ali and Yu, Qian and Avestimehr, A. Salman},
  journal={IEEE Transactions on Information Theory}, 
  title={A Fundamental Tradeoff Between Computation and Communication in Distributed Computing}, 
  year={2018},
  volume={64},
  number={1},
  pages={109-128},
  doi={10.1109/TIT.2017.2756959}}

@inproceedings{2017tandongradient,
  title={Gradient coding: Avoiding stragglers in distributed learning},
  author={Tandon, Rashish and Lei, Qi and Dimakis, Alexandros G and Karampatziakis, Nikos},
  booktitle={International Conference on Machine Learning},
  pages={3368--3376},
  year={2017},
  organization={PMLR}
}

@inproceedings{2019blockdesign,
  title={Gradient coding based on block designs for mitigating adversarial stragglers},
  author={Kadhe, Swanand and Koyluoglu, O Ozan and Ramchandran, Kannan},
  booktitle={2019 IEEE International Symposium on Information Theory (ISIT)},
  pages={2813--2817},
  year={2019}}

@article{WeiXuGe23,
  author={H. Wei and M. Xu and G. Ge},
  title={Robust Network Function Computation},
  journal={IEEE Transactions on Information Theory},
  year= 2023,
  volume= 69,
  number= 11,
  pages= {7070--7081}
}
\end{document}